\documentclass[12pt]{article}
\usepackage{amsfonts}
\usepackage[utf8]{inputenc}
\usepackage{bm}
\usepackage{amsmath}
\usepackage{epsfig}
\usepackage{enumitem}

\setlist[enumerate]{%
wide =0.5\parindent,
listparindent=0pt%
}

\newcommand{\bmat}{\left(\begin{array}}
\newcommand{\emat}{\end{array}\right)}

\def\gtrsim{\mathrel{\raise.3ex\hbox{$>$\kern-.75em\lower1ex\hbox{$\sim$}}}}

\def\a{\alpha}

\def\b{\beta}

\def\ov{\overline}
\def\un{\underline}

\def\-{\hphantom{-}}
\def\ov{\overline}
\def\s2{\frac{1}{\sqrt2}}

\def\mg{m_{3/2}}
\def\mg2{m^2_{3/2}}

\def\Dsl{\,\raise.15ex\hbox{/}\mkern-13.5mu D} 

\def\be{\begin{equation}}
\def\ee{\end{equation}}
\def\bea{\begin{eqnarray}}
\def\eea{\end{eqnarray}}

\newcommand{\nn}{\nonumber}

\begin{document}


\pagestyle{plain}

\makeatletter
\@addtoreset{equation}{section}
\makeatother
\renewcommand{\theequation}{\thesection.\arabic{equation}}
\pagestyle{empty}
\begin{center}
\ \

\vskip .5cm

\LARGE{\LARGE\bf $\alpha'$-corrections and their double formulation \\[10mm]}
\vskip 0.3cm

\large{Eric Lescano 
 \\[6mm]}

{\small  Instituto de Astronom\'ia y F\'isica del Espacio, Universidad de Buenos aires (UBA-CONICET)\\ [.01 cm]}
{\small\it Ciudad Universitaria, Pabell\'on IAFE, 1428 Buenos Aires, Argentina\\ [.3 cm]}

{\small \verb"elescano@iafe.uba.ar"}\\[1cm]

\small{\bf Abstract} \\[0.5cm]\end{center}
 
The present notes are based on three lectures, each ninety minutes long, prepared for the school
“Integrability, Dualities and Deformations", that ran from 23 to 27 August
2021 in Santiago de Compostela and virtually. These lectures, aimed at graduate students, require only a basic knowledge of string theory. The main goal is to introduce $\alpha'$-corrections to the gravitational sector of different formulations of closed string theory and to reformulate them using novel techniques based on double field theory.

\newpage

\setcounter{page}{1}
\pagestyle{plain}
\renewcommand{\thefootnote}{\arabic{footnote}}
\setcounter{footnote}{0}

\tableofcontents
\newpage

\section{Introduction}

In the last decades there has been a huge progress made in the study of the low energy limit of string theory. The corresponding field theories, collectively known as supergravity theories, describe particle dynamics and any string structure is completely suppressed. Many related programs, however, incorporate higher-order corrections beyond this low-energy limit and take some string features into account. Some current examples are: tests of duality conjectures \cite{Vafa}, microstate counting of black hole entropy \cite{SenBH}, moduli stabilization \cite{Grana}, the swampland program \cite{Palti}, among others.  
 
 Historically, $\alpha'$-corrections have been computed through scattering amplitudes \cite{GrossSloan} \cite{CN} \cite{MetsaevTseytlin} and beta-function computations \cite{Beta}. There are different string formulations and the order of the first non-trivial $\alpha'$-correction depends strongly on the theory. For instance, bosonic and heterotic string theories both contain four-derivative corrections in their leading-order Lagrangians ($\alpha'$ first-order correction) while type-II superstrings contain eight-derivative terms as their first corrections (corrections $\sim \alpha'^3$)\footnote{In these lectures we use the convention that $2i$-derivative terms in the string theory Lagrangian are of order $(\alpha')^{i-1}$.}. All these higher-derivative supergravities incorporate string effects to their dynamics.
 
 Other methods used to compute higher-order corrections are based on symmetry arguments: some terms of the effective string Lagrangian can be unambiguously fixed by means of symmetry transformations. For example, the classical Einstein-Hilbert Lagrangian is given by an object that transforms as a scalar under infinitesimal diffeomorphisms, the Ricci scalar. Therefore, a possible four-derivative correction to the Einstein-Hilbert action could be a $(\rm Riem)^2$ term, that also transforms as a scalar under infinitesimal diffeomorphisms. In \cite{BdR} $\alpha'$-corrections were studied using this method. The authors present a relation between gauge fields and the gravitational sector that allows to construct four-derivative corrections in a simple way. 
 
 In these lectures we will review both \cite{MetsaevTseytlin} and \cite{BdR} approaches, and we will generalize the construction of the latter in a double field theory (DFT) context \cite{DFT}. DFT is a novel framework to construct T-duality invariant theories, T-duality being an exact symmetry of string theory. Since DFT is invariant under the same symmetries as string theory, it is a promising framework to compute higher-derivative corrections. The idea of these lectures is to give a self-contained introduction to duality invariant approaches related to $\alpha'$-corrections and to present the state of the art of this topic.  

\subsection{Student background and string bibliography}

These lectures give an introduction to the study of $\alpha'$-corrections to the effective action of closed string theory. Our intention is to keep it self-contained, even though some knowledge on non-linear sigma models and their effective actions \cite{Cuerdas} is strongly recommended to get the most out of the course. A very pedagogical introduction to these topics can be found in \cite{Tong}. For people new to string theory, we recommend Barton Zwiebach's book \cite{Zwbook}.

\subsection{Outline}
The plan of the course is as follows:
\begin{itemize}
\item {\bf Lecture 1: Closed string theory and supergravity}

In part A we present the supergravity limit of closed string theory. We pay special attention to the NS-NS Lagrangian of type II superstring theory, since it is common to other closed string formulations. We analyze the symmetries of this Lagrangian, and the additional symmetries of the heterotic supergravity. We finish with a digression about the vielbein formalism.

In part B we present two approaches to study the four-derivative terms in closed string theory. We first present the Metsaev-Tseytlin approach \cite{MetsaevTseytlin}. The corresponding effective action is constructed taking into account three- and four-point scattering amplitudes for the massless states of the string. Then we present the Bergshoeff-de Roo approach \cite{BdR}, which is mostly based on supersymmetry considerations. We review an interesting relation between gauge and Lorentz vectors that allows to construct the gravitational $\alpha'$-corrections. We shall follow \cite{Tduality}, where the equivalence of these approaches was also discussed. 

\item{\bf Lecture 2: Double field theory and higher-derivative terms}

In part A we give a self-contained review of double field theory \cite{DFT}. We present the construction of the theory in four steps: We start by introducing the double geometry, describing its symmetries, discussing the fundamental fields and, finally, introducing the action principle. Then, we study the heterotic/gauged double field theory, and analyze the generalization of the previous steps for this case. We discuss the parametrizations needed in order to connect with the bosonic part of the heterotic supergravity.

In part B we introduce the generalized Green-Schwarz mechanism at the DFT level \cite{GreenDFT}. This mechanism consists of a suitable modification of the Lorentz transformation of the generalized frame \cite{Tduality}. We start by analyzing the heterotic DFT case and then we will generalize the ansatz for a bi-parametric case. We show that the parametrization of this deformation reproduces the Bergshoeff-de Roo approach after a suitable field redefinition of the vielbein. We finish this part with some comments about HSZ theory \cite{HSZ0}. This theory is not a string theory but it can be understood as a higher-derivative deformation of DFT. However the four-derivative correction of its supergravity Lagrangian is closely related to heterotic string theory while its six-derivative correction is related to bosonic string theory. 

\item{\bf Lecture 3: Extended space and higher-order formulations}

This lecture is focused on introducing a systematic procedure to finding higher-derivative corrections at the DFT level \cite{gbdr}. Our starting point is an $O(D,D+K)$ invariant theory (``gauged DFT''). By splitting the fields, metrics and parameters in terms of $O(D,D)$ degrees of freedom, we obtain a DFT formulation with an extra $O(D,D)$ vector that plays the role of gauge connection at the DFT level. We then identify this degree of freedom with a projection of the generalized flux. This method allows to construct the generalized Green-Schwarz mechanism \textit{and} the four-derivative action principle from geometric first principles. This identification between the $O(D,D)$ vector and the derivatives of the generalized frame works up to first-order.    

In part B we discuss about the extension of the previous method to higher-order. We present the generalized Bergshoeff-de Roo identification, which is an exact identification which allows us to construct symmetry transformations and action principles up to an arbitrary order. At the end, we present some current research lines around this generalization.

\end{itemize}

After each lecture a problem sheet is included. Those exercises with a * mark require a little bit more computations than the others. 

\newpage

\section{Lecture 1: Closed string theory and supergravity}
\label{Sec1}
String theory is a proposal to describe nature in a unified approach, where gravity and gauge symmetries coexist. The idea of string theory is very simple: to replace point-like particles with string excitations. Strings can be open or closed and their length are given by $l_{s}=\sqrt{\alpha'}$. As we learned from undergraduate courses, a string can oscillate in different modes, and the higher the mode, the higher the energy that is required. The lightest physical modes of the closed (super-)strings spectrum are massless and, in this sense, string theory predicts the existence of a finite number of massless particles and an infinite number of massive particles ($M^2\sim \frac{1}{\alpha'}$). From a phenomenological point of view it is common to focus only in the massless states, integrating out the massive ones. This procedure gives rise to a particular higher-derivative structure that corrects the low energy limit of the effective theory. The study of these $\alpha'$-corrections is the main topic of these lectures.  

An interesting aspect of string theory is that the starting point is not a 4-dimensional theory. The D-dimensional target space which contains our strings has an arbitrary dimension from a classical point of view, but a fixed and $D>4$ dimension if we want a consistent quantum theory \cite{Cuerdas}. This is one of the most exciting aspects of string theory and the reason is due to anomaly cancellation: the classical worldsheet formulation of the string is invariant under local Weyl transformations, and cannot be preserved in any dimension. 

In these lectures we are interested in three different formulations of closed string theories: the bosonic string with critical dimension $D=26$, the type II superstrings with critical dimension $D=10$, and the heterotic superstring also defined on a 10-dimensional target space. The superstring formulations are related to each other by string dualities, but each formulation receives different kinds of corrections. In order to avoid confusion, we will use the letter $D$ for the critical dimension of each string and we always consider $\frac{\sqrt{\alpha'}}{R}<<1$ with R the curvature of the target space. Thus, we are able to study the physics of the massless states of the different formulations by considering an $\alpha'$ expansion.

\subsection{Part A: low-energy effective formulations}
\label{A1}
\subsubsection{Action principles}
The low-energy effective action depends on the string formulation under consideration \cite{Vafa}. For example, the massless spectrum for bosonic string theory consists of a metric tensor $g_{\mu \nu}$, a 2-form or Kalb-Ramond field, $b_{\mu \nu}$, and a scalar field called the dilaton, $\phi$, where $\mu,\nu=0,\dots,D-1$ and in this case $D=26$. The action principle for this setup (in string frame) is given by
\bea
S = \int d^{D}x \sqrt{-g} e^{-2 \phi} (R + 4 \partial_{\mu} \phi \partial^{\mu}\phi - \frac{1}{12} H_{\mu \nu \rho} H^{\mu \nu \rho} + \mathcal{O}(\alpha')) \, ,
\label{bos0}
\eea
where $R$ is the Ricci scalar and 
\bea
H_{\mu \nu \rho} =  \partial_{\mu} b_{\nu \rho} + \partial_{\nu} b_{\rho \mu} + \partial_{\rho} b_{\mu \nu} \, ,
\eea
is a 3-form known as the $b$-field curvature. As the reader can easily verify, each term of the action principle (\ref{bos0}) contains two derivatives. The first-order correction (or the $\alpha'$-correction) to this effective action are given by four-derivative terms, while the second-order corrections are given by 6-derivative terms and so on. Since the order of the correction can be easily identified through the number of derivatives we adopt the convention $\alpha'=1$ to present a light notation. These $\alpha'$-corrections were historically computed from scattering amplitudes\footnote{The beta-function computation is a different way to obtain corrections. See \cite{Beta}.} as we will see in part B. Before moving to another formulation it is interesting to observe that the low-energy effective bosonic action consists of an Einstein-Hilbert action coupled to a 2-form and a scalar field.

As we have mentioned in the first part of this lecture, we are also interested in the type-II superstring formulation. There exist two different consistent frameworks, type IIA and type IIB. Both of them contain bosons and fermions in their spectrum and this is the main difference with respect to the bosonic string that only contains bosons. Schematically, the type-II effective Lagrangian can be written as,
\bea
L_{\rm{type-II}} = L_{\rm NS-NS} + L_{\rm R-R} + L_{\rm NS-R} + L_{\rm R-NS} \, .
\label{typeII0}
\eea
The first term, $L_{\rm NS-NS}$ , encodes the dynamics of the following massless fields: a metric tensor, a 2-form and a dilaton. Moreover, the two-derivative effective action principle for this part of the spectrum is exactly given by (\ref{bos0}), but considering $D=10$. The second term of (\ref{typeII0}), $L_{\rm R-R}$, describes the dynamics of a 1-form $C_{1}$ and a 3-form $C_{3}$ for type IIA superstring and a zero-form $C_{0}$, a two form $C_{2}$ and a four-form $C_{4}$ for type IIB superstring. Both $L_{\rm NS-NS}$ and $L_{\rm R-R}$ are bosonic Lagrangians, meaning that they depend on bosonic quantities. On the other hand, both $L_{\rm NS-R}$ and $L_{\rm R-NS}$ depend on fermionic degrees of freedom, related to the bosonic spectrum through ${\cal N}=2$ supersymmetry. This symmetry interchanges bosons and fermions. The critical dimension of this formulation is $D=10$. The Lagrangian (\ref{typeII0}) is our first example of a supergravity Lagrangian, since it contains both a dynamical metric tensor and fermionic degrees of freedom related by supersymmetry. Indeed, the low energy limit of all superstring theory formulations is given by a supergravity. Moreover, supergravity theories in arbitrary dimensions can be studied independently of string theory. It turns out that the first $\alpha'$-corrections for the type II superstrings consist of eight-derivative terms, so neither $\alpha'$ nor $\alpha'^2$ corrections are present in this theory. 

Last but not least, we consider heterotic string theory \cite{gw}. This string is a hybrid formulation. To construct it we consider a left-right decomposition for the closed string oscillation. This is analogous to a travelling wave decomposition, where left and right are the propagation directions of a wave on the closed string. The heterotic string contains a $D=26$ bosonic string formulation on its left part, compatified on a $16$-dimensional torus (for dimensional consistency), and a $D=10$ superstring formulation on its right side. The 10-dimensional bosonic and massless degrees of freedom are a metric tensor, a 2-form, a dilaton and a non-Abelian gauge field $A_{\mu i}$. The index $i$ is a gauge index that runs from $1,\dots,n$ with $n$ the dimension of the gauge group, $E_{8}\times E_{8}$ or $SO(32)$, depending on the heterotic formulation.

This superstring theory is invariant under ${\cal N}=1$ supersymmetry so it contains fermionic degrees of freedom besides the former fields. The effective action is given by a heterotic supergravity,      
\bea
S_{\rm{het}} = \int d^{D}x \sqrt{-g} e^{-2 \phi} (R + 4 \partial_{\mu} \phi \partial^{\mu}\phi - \frac{1}{12} \bar{H}_{\mu \nu \rho} \bar{H}^{\mu \nu \rho} - \frac14 F_{\mu \nu}{}^{i} F^{\mu \nu}{}_{i} + L_{f} + \mathcal{O}(\alpha')) \, ,
\label{het0}
\eea
where $L_{f}$ are two-derivative fermionic terms and\footnote{We use the convention $T_{[\mu \nu]} = \frac12(T_{\mu \nu} - T_{\nu \mu})$ and $T_{(\mu \nu)} = \frac12(T_{\mu \nu} + T_{\nu \mu})$.}
\bea
F_{\mu \nu}{}^{i} = 2 \partial_{[\mu} A_{\nu]}{}^{i} - f^{i}{}_{jk} A_{\mu}{}^{j} A_{\nu}{}^{k} \, ,
\eea
is the gauge field curvature. In the previous expression gauge indices are contracted using a constant Cartan-Killing metric $\kappa_{i j}$ with inverse $\kappa^{i j}$. The first higher-derivative correction to this action principle is a correction of order $\alpha'$. We will discuss the form of these terms in the next section. In order to keep track of the $\alpha'$ terms, we assume that the structure constants $f_{ijk}$ count as a derivative. So, the first non-trivial $\alpha'$-corrections are terms with $f^{m} \partial^{n}$ such that $n+m=4$, and so on. Therefore we still think in (\ref{het0}) as a zeroth-order Lagrangian.  

An important difference with respect to the previous formulations is that the curvature of the $b$-field contains extra terms \cite{Green},
\be
\bar H_{\mu\nu\rho}=3\left(\partial_{[\mu}b_{\nu\rho]}-C_{\mu\nu\rho}^{(g)}\right)\, , \label{barH}
\ee 
where $C_{\mu\nu\rho}^{(g)}$ is a so-called Chern-Simons 3-form, defined as
\be
C_{\mu\nu\rho}^{(g)}=A^i_{[\mu}\partial_\nu A_{\rho]i}-\frac13 f_{ijk}A_\mu^i A_\nu^jA_\rho^k \, . 
\ee
The necessity for this additional contribution will be clear when we inspect the symmetry rules of this formulation.  

\subsubsection{Symmetry transformations in bosonic/type II supergravity}
We start by discussing the symmetry transformations of the universal sector,
\bea
S = \int d^{D}x \sqrt{-g} e^{-2 \phi} (R + 4 \partial_{\mu} \phi \partial^{\mu}\phi - \frac{1}{12} H_{\mu \nu \rho} H^{\mu \nu \rho}) \, .
\label{univ}
\eea
This action is common to all the effective close string formulations, and it is invariant under:
\begin{itemize}
\item {\bf Infinitesimal diffeomorphisms:} When acting on an arbitrary vector $v^{\mu}$ of weight $\omega$ this transformation takes the form
\bea
\delta_{\xi} v^{\mu} = L_{\xi} v^{\mu} = \xi^{\nu} (\partial_{\nu} v^{\mu}) - (\partial_{\nu} \xi^{\mu}) v^{\nu} + \omega (\partial_{\nu} \xi^{\nu}) v^{\mu}\, ,
\eea
where $\xi^{\mu}$ is an arbitrary parameter and $L_{\xi}$ is the Lie derivative. The fundamental fields transform as \footnote{We use the convention that contracted indices are not part of the (anti-)symmetrizers.}
\bea
\delta_{\xi} g_{\mu \nu} & = & L_{\xi} g_{\mu \nu} = \xi^{\rho} (\partial_{\rho} g_{\mu \nu}) + 2(\partial_{(\mu} \xi^{\rho}) g_{\rho \nu)} \, , \\
\delta_{\xi} b_{\mu \nu} & = & L_{\xi} b_{\mu \nu} = \xi^{\rho} (\partial_{\rho} b_{\mu \nu}) + 2(\partial_{[\mu} \xi^{\rho}) b_{\rho \nu]} \, , \\
\delta_{\xi} \phi & = & L_{\xi} \phi = \xi^{\rho} (\partial_{\rho} \phi) \, ,
\eea
and the closure of these transformations acting on a generic (1,1) tensor $T_{\mu}{}^{\nu}$, \textit{i.e.},
\bea
\Big[\delta_{\xi_1},\delta_{\xi_2} \Big] T_{\mu}{}^{\nu} = \delta_{\xi_{21}} T_{\mu}{}^{\nu}
\eea
is given by the Lie bracket 
\bea
\xi^{\mu}_{12}(x) = \xi^{\rho}_{1} \frac{\partial \xi^{\mu}_{2}}{\partial x^{\rho}} - (1 \leftrightarrow 2)  \, .
\label{LieBra}
\eea
At this point we observe that the transformation of the derivative of a vector, $\delta_{\xi} (\partial_{\mu} v^{\nu}) =  \partial_{\mu}(\delta_{\xi} v^{\nu})$, does not transform as a (1,1) tensor. Consequently, we define a covariant derivative in the usual way,
\bea
\nabla_{\rho} T_{\mu}{}^{\nu} = \partial_{\rho} T_{\mu}{}^{\nu} - \Gamma_{\rho \mu}^{\sigma} T_{\sigma}{}^{\nu} + \Gamma_{\rho \sigma}^{\nu} T_{\mu}{}^{\sigma} \, ,
\label{nabla}
\eea
where $\Gamma_{\mu \nu}^{\rho}$ is the Christoffel symbol, $\Gamma_{\mu \nu}^{\rho} = \frac12 g^{\rho \sigma} (\partial_{\mu} g_{\nu \sigma} + \partial_{\nu} g_{\mu \sigma} - \partial_{\sigma} g_{\mu \nu})$.

Each term of the Lagrangian
\be
L = e^{-2\phi}(R + 4 \partial_\mu \phi \partial^{\mu}\phi - \frac{1}{12} H_{\mu \nu \lambda} H^{\mu \nu \lambda})
\label{Lag}
\ee
transforms as a scalar (with $\omega=0$),
\be
\delta_{\xi} L = \xi^{\mu} \partial_{\mu} L \, , 
\ee
while the square root of the determinant transforms as
\be
\delta_{\xi}(\sqrt{-g}) = \partial_{\mu}(\sqrt{-g} \xi^{\mu}) \, .  
\ee
Consequently, the universal action (\ref{univ})
is invariant up to total derivatives.

\item {\bf Abelian gauge transformations:} This transformation only acts on the $b$-field due to its 2-form nature,
\be
\delta_{\zeta} b_{\mu \nu} = 2 \partial_{[\mu} \zeta_{\nu]} \, ,
\ee
where $\zeta$ is an arbitrary parameter. Since 
\bea
\delta_{\zeta} H_{\mu \nu \rho} = 3 \partial_{[\mu}(\delta b_{\nu \rho]}) = 6 \partial_{[\mu}(\partial_{\nu} \zeta_{\rho]}) = 0 \, ,
\eea
the Lagrangian (\ref{Lag}) is gauge invariant.

\subsubsection{Symmetry transformations in heterotic supergravity}

The low-energy heterotic effective Lagrangian,
\be
L_{\rm{het}} = e^{-2\phi}(R + 4 \partial_\mu \phi \partial^{\mu}\phi - \frac{1}{12} \bar H_{\mu \nu \lambda} \bar H^{\mu \nu \lambda} - \frac14 F_{\mu \nu}{}^i F^{\mu \nu}{}_{i}) \, ,
\label{LagHet}
\ee
is invariant under the symmetries previously described since $A_{\mu i}$ transforms as a 1-form under infinitesimal diffeomorphisms, \textit{i.e.},
\bea
\delta_{\xi} A_{\mu i} = L_{\xi} A_{\mu i} = \xi^{\rho} (\partial_{\rho} A_{\mu i}) + (\partial_{\mu} \xi^{\rho}) A_{\rho i} \, ,
\eea
and is invariant under Abelian gauge transformations. The Lagrangian (\ref{LagHet}) is also invariant under non-Abelian gauge transformations.

\item{\bf Non-Abelian gauge transformations:}
The non-Abelian gauge transformations act on $A_{\mu i}$ and $b_{\mu \nu}$ in the following way,
\bea
\delta_{\lambda}A_\mu^i & = & \partial_\mu \lambda^i+f^i{}_{jk}\lambda^jA_\mu^k\, , \\ \delta_{\lambda} b_{\mu\nu} & = & -(\partial_{[\mu} \lambda^i) A_{\nu]i}\, , \label{gauge0}
\eea
where $\lambda^{i}$ is an arbitrary parameter. Considering an arbitrary gauge vector $v^{i}$, the partial derivative of this vector is not covariant and we need to define a covariant derivative as
\bea
\nabla_{\mu} v^{i} = \partial_{\mu} v^{i} - f^{i}{}_{jk} A_{\mu}{}^{j} v^{k} \, .
\eea 
Here we use the same notation $\nabla$ for the gauge covariant derivative as in (\ref{nabla}), so our convention is that $\nabla$ covariantizes the derivative of an object with respect to all the symmetries that the object transforms under. On the other hand,  $F_{\mu \nu i}$ transforms covariantly under non-Abelian gauge transformations
\bea
\delta_{\lambda} F_{\mu \nu i} = f_{ijk} \lambda^{j} F_{\mu \nu}{}^{k} \, , 
\eea
unlike $A_{\mu i}$ whose transformation is not covariant due to the presence of the $\partial_{\mu}\lambda^i$ term. Moreover, we need $\delta \bar H_{\mu \nu \rho}=0$ to ensure the non-Abelian gauge invariance of the effective action. Now it is clear the Chern-Simons terms in (\ref{barH}) were need as they transform as
\bea
\delta_{\lambda} C^{(g)}_{\mu \nu \rho} = \partial_{[\mu}(A_{\nu}{}^{i} \partial_{\rho]} \lambda_{i}) \, .
\eea
The previous transformation exactly cancels the $3 \partial_{[\mu}(\delta_{\lambda} b_{\nu \rho]})$ contribution such that (\ref{het0}) is non-Abelian gauge invariant. 
\subsubsection{Digression: vielbein formalism}

All supersymmetric invariant theories, such as type II or heterotic string theory, must be written in the vielbein formalism of general relativity, instead of the metric formalism, in order to couple the fermions. Consequently, we consider flat vectors $v^a$ defined on the tangent space of each point of the target space, where the indices $a,b=0,\dots,D-1$ are known as ``flat indices''. Flat vectors transform under Lorentz transformations according to
\bea
\delta_{\Lambda} v^{a} = v^{b} \Lambda_{b}{}^{a}
\eea
where all the contractions are made with a constant flat (inverse) metric $\eta^{ab}$. The Lorentz parameter satisfies $\Lambda_{ab}= - \Lambda_{ba}$. Since we are abandoning the metric formulation, we need to consider a new fundamental field $e_{\mu}{}^{a}$, the vielbein, whose inverse is given by $e^{\mu}{}_{a}$. This object satisfies,
\bea
e_{\mu}{}^{a} \eta_{ab} e_{\nu}{}^{b} = g_{\mu \nu} \, ,
\eea
and transforms covariantly under infinitesimal diffeomorphisms and Lorentz transformations,
\bea
\delta_{\xi,\Lambda} e_{\mu}{}^{a} & = & \xi^{\nu} \partial_{\nu} e_{\mu}{}^{a} + \partial_{\mu} \xi^{\nu} e_{\nu}{}^{a} + e_{\mu}{}^{b} \Lambda_{b}{}^{a} \, , \\ 
\delta_{\xi,\Lambda} e^{\mu}{}_{a} & = &  \xi^{\nu}\partial_{\nu}  e^{\mu}{}_{a}  - \partial_{\nu}\xi^{\mu} e^{\nu}{}_{a} + e^{\mu}{}_{b} \Lambda^{b}{}_{a} \, .
\label{inverse0}
\eea
The other fundamental fields, $b_{\mu \nu}, A_{\mu i}$ and $\phi$ are Lorentz invariant. Analogously to what happens with infinitesimal diffeomorphisms or non-Abelian gauge transformations, the transformation of the partial derivative of a flat vector does not match with the transformation of a flat tensor. Considering a generic flat vector $v^a$, we define a flat covariant derivative as
\bea
\nabla_{\mu} v^{a} = \partial_{\mu} v^{a} - w_{\mu}{}^{a}{}_{b} v^{b} \, ,
\eea
where $w_{\mu ab}$ is the spin connection. Imposing
\bea
\nabla_{\mu} e_{\nu}{}^{a} = \partial_{\mu} e_{\nu}{}^{a} - \Gamma_{\mu \nu}^{\rho} e_{\rho}{}^{a} - w_{\mu}{}^{a}{}_{b} e_{\nu}{}^{b} = 0 \, ,
\label{compae}
\eea
we can fully determine the spin connection in terms of the vielbein, $w_{\mu b c} = w_{\mu b c}(e)$. The 2-form version of the Riemann tensor can be written in terms of the vielbein as
\bea
R_{\mu \nu a b} = - 2 \partial_{[\mu} w_{\nu] a b} + 2 w_{[\mu a}{}^{c} w_{\nu] c b} \, ,
\eea
and the previous object satisfies $R_{\mu \nu a b} e^{\rho a} e_{\sigma}{}^{b} = R^{\rho}{}_{\sigma \mu \nu}$
where 
\bea
R^{\rho}{}_{\sigma \mu \nu} = 2 \partial_{[\mu} \Gamma^{\rho}_{\nu] \sigma} + 2 \Gamma^{\rho}_{[\mu \tau} \Gamma_{\nu] \sigma}^{\tau} \, 
\eea
is the Riemann tensor in terms of the metric. Traces of the Riemann tensor give the Ricci tensor and scalar, respectively
\bea
R_{\mu \nu} = R^{\rho}{}_{\mu \rho \nu} \, , \quad R = R_{\mu \nu} g^{\mu \nu} \ .
\eea

\end{itemize}
\subsection{Part B: $\alpha'$-corrections}
\label{B1}
In this part we will study two possible ways of presenting the gravitational $\alpha'$-corrections. These contributions are four-derivative terms that correct the low energy effective Lagrangian of bosonic/heterotic string theory. On the one hand, we will study the Metsaev-Tseytlin approach, which is obtained considering scattering amplitudes. Then, we will study the Bergshoeff-de Roo approach, related to the former using field redefinitions.

\subsubsection{The Metsaev-Tseytlin approach}

The four-derivative corrections to the universal action principle of the different formulations of closed string theory were historically computed considering three- and four-point scattering amplitudes for the massless states \cite{GrossSloan} \cite{CN} \cite{MetsaevTseytlin}. This method is based on the study of the different string interactions through the S-matrix, in order to construct an effective Lagrangian that be able to reproduce that interactions. The effective action, originally computed by Metsaev and Tseytlin, takes the form
\bea
S_{MT} = \int d^Dx \sqrt{-g} e^{-2\phi} (L^{(0)} + L^{(1)}_{MT}) \, ,
\label{fullA}
\eea
where $L^{(0)}$ is given by (\ref{bos0}) and
\bea
L^{(1)}_{\rm MT} & = & - \frac{a+b}{8} \Big[  R_{\mu \nu \rho \sigma} R^{\mu \nu \rho \sigma} - \frac12 H^{\mu \nu \rho} H_{\mu \sigma \lambda} R_{\nu \rho}{}^{\sigma \lambda} + \frac{1}{24} H^4 - \frac18 H^2_{\mu \nu} H^{2 \mu \nu} \Big] \nn \\ && + \frac{a-b}{4} H^{\mu \nu \rho} C_{\mu \nu \rho} \, 
\label{MT}
\eea
corresponds to the first-order $\alpha'$-correction with $a$ and $b$ units of $\alpha'$. In the previous expression we have introduced the following quantities,
\bea
\label{CSdef}
C_{\mu \nu \rho} & = & w_{[\mu}{}^{ab} \partial_{\nu} w_{\rho]ab} + \frac23 w_{[\mu}{}^{ab} w_{\nu b}{}^{c} w_{\rho] ca} \, , \\
H^2_{\mu \nu} & = & H_{\mu}{}^{\rho \sigma} H_{\nu \rho \sigma} \, , \\
H^2 & = & H_{\mu \nu \rho} H^{\mu \nu \rho} \, .
\eea
The bi-parametric form of the Lagrangian (\ref{MT}) was introduced in \cite{Tduality}. As we will see, this compact form allows comparisons between different formulations. In order to find $\a'$-corrections we need to impose
 \bea
 \label{cases}
 (a,b) = 
   \begin{cases} 
      (-1,-1)  & \mbox{bosonic string \, , }  \\
      (-1,0)  & \mbox{heterotic string \, ,} \\ 
      (0,0) & \mbox{type II strings \, .} 
   \end{cases}
\eea
At this point it is important to recall that we are showing only the gravitational corrections to the universal NS-NS sector which contains only a metric tensor, a 2-form and a dilaton which means that, for instance, we are turning off the non-Abelian gauge sector of the heterotic supergravity. 

The full action (\ref{fullA}) requires a modification of the Lorentz transformation of the $b$-field at order $\alpha'$ \footnote{A zeroth-order Lorentz transformation has no derivatives, so a first-order transformation must contain two derivatives.},
\bea
\delta_{\Lambda} b_{\mu \nu} = \frac12 (a-b) \partial_{[\mu} \Lambda^{a b} w_{\nu] a b} \, .
\eea
This expression is the so-called Green-Schwarz transformation \cite{Green} and is a cornerstone of  string theory phenomenology. We can absorb the second line of (\ref{MT}) by deforming the curvature of the $b$-field once again
\bea
\bar H_{\mu \nu \rho} = 3 \Big(\partial_{[\mu} b_{\nu \rho]} + \frac12 (b-a) C_{\mu \nu \rho}\Big) \, .
\eea
As the reader can easily check, the Green-Schwarz transformation and the 3-form deformation are present only in the heterotic case. Moreover, the new terms in the curvature of the $b$-field guarantees that $\bar H_{\mu \nu \rho}$ transforms trivially under Lorentz transformations,
\bea
\delta_{\Lambda} \bar H_{\mu \nu \rho} = 0
\eea
and then $L_{\rm MT}$ is Lorentz invariant when four-derivative corrections are taking into account.

\subsubsection{The Bergshoeff-de Roo approach}
Another historical way to compute higher-order derivative corrections is to appeal to symmetry arguments. For instance, if we focus on the bosonic terms in the heterotic supergravity Lagrangian,
\be
L_{het} = e^{-2\phi}(R + 4 \partial_\mu \phi \partial^{\mu}\phi - \frac{1}{12} \bar H_{\mu \nu \lambda} \bar H^{\mu \nu \lambda} - \frac14 F_{\mu \nu}{}^i F^{\mu \nu}{}_{i}) \, ,
\label{NS}
\ee
each term of this Lagrangian is invariant under infinitesimal diffeomorphisms, Abelian/non-Abelian gauge transformations and Lorentz transformations. In this sense, the coefficients $\Big\{1,4,-\frac{1}{12},-\frac14 \Big\}$ are not determined by these symmetries. However,  ${\cal N}=1$ supersymmetric invariance unequivocally determines all the coefficients. In other words, if we change one of these coefficients by hand the full action is not ${\cal N}=1$ supersymmetric invariant anymore.

Using this kind of argument, Bergshoeff and de Roo obtained the following corrections \cite{BdR}, 
\bea
S_{\rm BdR} = \int d^Dx \sqrt{-g} e^{-2\phi} (\tilde L_{0} + L_{1})
\label{BdR}
\eea
where
\bea
L_{1} = - \frac{a}{8} R^{(-)}_{\mu \nu}{}^{ab} R^{(-)\mu \nu}{}_{ab} - \frac{b}{8} R^{(+)}_{\mu \nu}{}^{ab} R^{(+)\mu \nu}{}_{ab}
\eea
and the $(\pm)$ notation means that we need to add a torsion term in the spin connection $w_{\mu ab} \rightarrow w^{(\pm)}_{\mu a b} = w_{\mu a b} \pm \frac12 H_{\mu \nu \rho} e^{\nu}{}_{a} e^{\rho}{}_{b}$. The zeroth-order Lagrangian is given by
\bea
\tilde L_{0} = R + 4 \partial_{\mu}\phi \partial^{\mu} \phi - \frac{1}{12} \tilde H^2 \, ,
\eea
where $\tilde H_{\mu \nu \rho} = H_{\mu \nu \rho} - \frac32 a C^{(-)}_{\mu \nu \rho} + \frac32 b C^{(+)}_{\mu \nu \rho} \, $ and $C_{\mu \nu \rho}$ was defined in (\ref{CSdef}). In order to reproduce the different corrections for each string we need to use (\ref{cases}) again. This compact form of the Bergshoeff-de Roo action was introduced in \cite{Tduality}.

The action principle (\ref{BdR}) is invariant under Lorentz transformations if we impose the following Green-Schwarz mechanism,
\bea
\delta_{\Lambda} b_{\mu \nu} = \frac12 (a-b) \partial_{[\mu} \Lambda^{ab} w_{\nu]ab} - \frac14 (a+b) \partial_{[\mu} \Lambda^{ab} H_{\nu]ab} \, .
\eea
To prove the equivalence between $S_{\rm MT}$ and $S_{\rm BdR}$ we need to consider the following non-covariant field redefinition for the $b$-field,
\bea
b^{\rm{MT}}_{\mu \nu} = b^{\rm{BdR}}_{\mu \nu} + \frac14 (a+b) H_{[\mu}{}^{ab} w_{\nu]ab} \, ,
\label{bequivalence}
\eea
and extra covariant field redefinitions \cite{Tduality}. Since both approaches are equivalent, let us now fix the parameters $(a,b)=(-1,0)$ in (\ref{BdR}) in order to review an interesting relation found by Bergshoeff-de Roo. The action principle is given by
\bea
\int d^Dx \sqrt{-g} e^{-2\phi} (R + 4 \partial_\mu \phi \partial^{\mu}\phi - \frac{1}{12} \tilde H_{\mu \nu \lambda} \tilde H^{\mu \nu \lambda} + \frac18 R^{(-)}_{\mu \nu}{}^{ab} R^{(-)\mu \nu}{}_{ab})
\label{LagHetL}
\eea
where $\tilde H_{\mu \nu \rho}= 3\left(\partial_{[\mu}b_{\nu\rho]}+\frac12 C^{(-)}_{\mu\nu\rho}\right)\, $ and the only higher-derivative correction to the symmetry transformations is the Green-Schwarz mechanism,
\bea
\delta_{\Lambda} b_{\mu \nu} = - \frac12 \partial_{[\mu} \Lambda^{ab} w^{(-)}_{\nu]ab} \, . 
\label{Lorentz1}
\eea

The key observation here is that the structure of the action and transformations resembles the form of a Yang-Mills theory coupled to the NS-NS sector. In fact, the Lagrangian (\ref{LagHetL}) is rather similar to (\ref{LagHet}), and the Green-Schwarz transformation (\ref{Lorentz1}) is the analogous to (\ref{gauge0}). Moreover, in both Lagrangians the 3-form is deformed by Chern-Simons terms. Therefore, here we review an alternative way for constructing the four-derivative gravitational corrections in the heterotic supergravity. Let us define the gauge generators $(t_{i})^{ab}$ such that
\bea
(t_{i})^{ab} (t^{i})_{cd} & = & - \delta_{[c}^{a} \delta_{d]}^{b} \, , \\
(t^{i})_{ab} (t_{j})^{ab} & = & - \delta_{j}^i \, , \\
f_{i}{}^{kl} (t^i)_{ab} & = & 2 \sqrt{2} (t^k)_{[a|c} (t^l)^{c}{}_{b]} \, .
\label{map}
\eea
The gauge generators are antisymmetric in their flat indices, $(t_{i})^{ab}=-(t_{i})^{ba}$ and, thanks to them, each generic gauge vector/parameter can be written in the following way,
\bea
v^{i} & = & - v^{ab} (t^i)_{ab} \, , \\
\lambda^i & = & - \lambda^{ab} (t^i)_{ab} \, .
\eea
If we apply the previous relations to $\delta_{\lambda}b_{\mu \nu}$ in (\ref{gauge0}) we find
\bea
\delta_{\lambda} b_{\mu \nu} = \partial_{[\mu} \lambda_{a b} A_{\nu]}{}^{a b} \, .
\eea
Now we identify 
\bea
\label{sugraidl}
\lambda_{ab} & = & - \frac{1}{\sqrt 2}\Lambda_{ab} \, , \\ A_{\nu}{}^{a b} & = & \frac{1}{\sqrt 2} w^{(-)}_{\nu}{}^{ab} \, ,
\label{sugraidA}
\eea
obtaining the Lorentz Green-Schwarz transformation from the gauge Green-Schwarz transformation $\delta_{\lambda} b_{\mu \nu} \rightarrow \delta_{\Lambda} b_{\mu \nu}$.
Moreover if we apply (\ref{map}) at the level of the action considering the identification (\ref{sugraidA}), it is straightforward to find
\bea
\label{Rrel}
-\frac{1}{4} F_{\mu \nu}{}^{i} F^{\mu \nu}{}_{i} & \rightarrow & \frac{1}{8} R^{(-)}_{\mu \nu}{}^{ab} R^{(-)\mu \nu}{}_{ab} \\ \label{Crel}
3(\partial_{[\mu} b_{\nu \rho]} - C^{(g)}_{\mu \nu \rho}) & \rightarrow & 3 (\partial_{[\mu} b_{\nu \rho]} + \frac12 C^{(-)}_{\mu \nu \rho}) \, .
\eea
The assignment (\ref{Rrel}) indicates that we can identify,
\bea
F_{\mu \nu}{}^{ab} = - \frac{1}{\sqrt 2} R^{(-)}_{\mu \nu}{}^{ab} \, .
\eea
In \cite{BdR}, the authors use the previous identification to construct a higher-derivative heterotic supergravity. Nowadays, an active topic of investigation is to follow a similar program by considering DFT. The main goal of the next lectures is to show that it is possible to perform identifications such as (\ref{sugraidl}) and (\ref{sugraidA}) in a DFT framework to find a higher-derivative DFT formulation related to the Bergshoeff-de Roo approach.
\newpage
\subsection{Exercises}
\begin{enumerate}
    \item Show that the curvature for the $b$-field can be written as $H_{\mu \nu \rho}=\nabla_{[\mu} b_{\nu \rho]}$ in bosonic supergravity. Then show its Bianchi identity $\nabla_{[\mu} H_{\nu \rho \sigma]}=0$.
    
    \item Consider a generic vector $v^{\mu}$ and show that the closure $\Big[\delta_{\xi_1},\delta_{\xi_2} \Big] v^{\mu} = \delta_{\xi_{21}} v^{\mu} \, $ holds if $\xi_{12}=-\xi_{21}$ is given by (\ref{LieBra}).

\item Using the Jacobi identity for $f_{ijk}$, \textit{i.e.}, $
f_{il}{}^{m} f_{jk}{}^{l} + f_{jl}{}^{m} f_{ki}{}^{l} + f_{kl}{}^{m} f_{ij}{}^{l} = 0 \, , $ show that $F_{\mu \nu i}$ transforms covariantly under non-Abelian gauge transformations.

\item Using $f^{i}{}_{jk} f_{i}{}^{mn} = 2 \alpha' \delta_{[j}^{m} \delta_{k]}^{n}$ compute $\delta_{\lambda} C^{(g)}_{\mu \nu \rho}$.

\item Consider a generic gauge vector $v^{i}$. Show $\big[\nabla_{\mu},\nabla_{\nu} \big] v^{i} = f_{jk}{}^{i} v^{j} F_{\mu \nu}{}^{k}$.

\item Show that $\nabla_{[\mu} C^{(g)}_{\nu \rho \sigma]}=\frac14 F_{[\mu \nu}{}^{i} F_{\rho \sigma]i}$. Then use this result to obtain the Bianchi identity for $\bar H_{\mu \nu \rho}$.

\item Consider a generic flat vector $v^a$ and compute $\delta_{\Lambda}(\partial_{\mu} v^{a})$ and $\delta_{\Lambda}(\nabla_{\mu} v^{a})$. Compare both expressions to obtain $\delta_{\Lambda} w_{\mu a b}$.

\item Use the vielbein compatibility to obtain $w_{abc} = -e^\mu{}_{[a}{} e^\nu{}_{b]}{}\partial_\mu e_{\nu c} +e^\mu{}_{[a}{} e^\nu{}_{c]}{}\partial_\mu e_{\nu b}
+e^\mu{}_{[b}{} e^\nu{}_{c]}{}\partial_\mu e_{\nu a}$ \, .

\item Show that $\delta_{\Lambda} \bar H_{\mu \nu \rho}=0$ in the Metsaev-Tseytlin approach and $\delta_{\Lambda} \tilde H_{\mu \nu \rho}=0$ in the Bergshoeff-de Roo approach.

\item Compute $\delta_{\Lambda} b^{MT}_{\mu \nu} $ from (\ref{bequivalence}) to prove that the b-redefinition holds.

\item Compute the following decomposition $R^{(\pm)}_{\mu \nu ab} = R_{\mu \nu a b} \mp \nabla_{[\mu} H_{\nu]ab} + \frac12 H_{[\mu a}{}^{c} H_{\nu] c b}$ where $H_{\mu a b} = H_{\mu \nu \rho} e^{\nu}{}_{a} e^{\rho}{}_{b}$.

\item Show the following identities: 
\bea
(\nabla_{\mu} H_{\nu \rho \sigma}) H^{\mu \rho}{}_{\tau} H^{\nu \sigma \tau} = 0 \, , \quad (\nabla_{\mu} H^{\nu \rho \sigma}) R^{\mu}{}_{\nu \rho \sigma} = 0 \, .
\eea

\item Show that the assignments (\ref{Rrel}) and (\ref{Crel}) hold.

\end{enumerate}

\section{Lecture 2: Double field theory and higher-derivative terms}
\label{Sec2}

As we mentioned in the previous lecture, the different formulations of string theory are defined on a $D$-dimensional manifold. From a phenomenological point of view, this $D$-dimensional target space is often divided
into an external non-compact space-time $M_{D-N}$ and an internal compact space $M_{N}$,
\bea
M_{D} = M_{D-N} \times M_{N} \, ,
\eea
where the simplest case is to consider an internal toroidal manifold. In this scenario, the resulting theory is invariant under different symmetries. One of these symmetries is T-duality or, more precisely, $O(N,N)$ invariance. This symmetry group appears after compactifying the universal NS-NS sector on a N-dimensional torus and it is an exact symmetry of string theory \cite{Ashoke}. 

The main idea of double field theory (DFT) \cite{DFT} \cite{Siegel}  \cite{DFTextra} is to rewrite a supergravity theory as a T-duality invariant theory \textit{before} compactification. In this sense, DFT represents an $O(D,D)$ invariant theory, where $D$ is the dimension of the target space of the embedded supergravity\footnote{For complementary material about DFT we recommend \cite{ReviewDFT}.}. All the DFT fields and parameters are $O(D,D)$ multiplets or, in other words, covariant objects. DFT coordinates lie in the fundamental representation of $O(D,D)$ whose dimension is $2D$. Then $X^{M}=(\tilde{x}_{\mu},x^{\mu})$ are coordinates of a double space. Here $x^{\mu}$ are the coordinates of the embedded supergravity, and $\tilde{x}_{\mu}$ are $D$ extra coordinates that we need for consistency. The dual coordinates are taken away considering that fields and parameters depend only on $x^{\mu}$ and therefore $\tilde \partial=0$. This is, in fact, the simplest solution to the ``strong constraint'',
\be
\partial_{M} (\partial^{M} \star) = (\partial_{M} \star) (\partial^{M} \star) = 0 \, ,
\label{SC}
\ee
where $\star$ means a product of arbitrary DFT fields/parameters. Contractions here are given by the $O(D,D)$ invariant metric, $\eta_{M N}$. 

\subsection{Part A: zeroth-order formulation}
\label{A2}

The DFT construction can be performed following four steps, 
\begin{enumerate}
\item {\bf Double geometry}: We consider a double geometry with coordinates $X^{M}$ with $M=0,\dots,2D-1$. We equip a group invariant metric $\eta_{MN}$ and its inverse $\eta^{MN}$. These metrics are used to lower and raise double curved indices.  On each point of the double space we consider a double tangent space, so we are able to define flat vectors $V^A$, with $A$ a flat index, $A=0,\dots,2D-1$. Then we consider two additional invariant and flat metrics $\eta_{A B}$ and ${\cal H}_{A B}$. The former is used to lower flat indices and both of them are used to construct the following flat projectors,
\bea
P_{AB} = \frac{1}{2}\left(\eta_{AB} - {\cal H}_{AB}\right) \, , \ \  \ \
\ov{P}_{AB} = \frac{1}{2}\left(\eta_{AB} + {\cal H}_{AB}\right)\ \, ,
\eea
which satisfy
\bea
&{\overline{P}}_{{A B}} {\overline{P}}^{ B}{}_{C}={\overline{P}}_{{A C}}\, , &\quad {P}_{{A B}} {P}^{B}{}_{C}={P}_{{A C}}, \nn\\
&{P}_{{A  B}}{\overline{P}}^{B}{}_{ C} = {\overline{P}}_{ {A B}}  {P}^{ B}{}_{C} = 0\, ,  &\quad {\overline{P}}_{{AB}} + {P}_{{A B}} = \eta_{{A B}}\,.
\eea
Thus, $P$ and $\bar P$ project onto complementary orthogonal subspaces and we can write any arbitrary vector as
\bea
V^{A} = P^{A}{}_{B} V^{B} + \ov P^{A}{}_{B} V^{B} = V^{\un A} + V^{\ov A} \, .
\label{projflat}
\eea
The notation underline and overline means that we use the projectors to lower/raise the indices, instead of using an invariant metric.

\item {\bf Symmetries:} DFT is a T-duality invariant formulation. Moreover, all the fields and parameters are written in representations of the duality group and, consequently, duality invariance is always guaranteed. 
Infinitesimal $O(D,D)$ transformations acting on an arbitrary double vector reads
\bea
\delta_h V^M = V^{N} h_{N}{}^{M} \, ,
\label{duality}
\eea
where $h \in o(D, D)$ \footnote{We use the lower-case because $h$ belongs to the associated algebra.} is an arbitrary parameter. 

We can also define generalized diffeomorphisms. These are infinitesimal transformations acting on a generic double vector $V^M$ through a generalized Lie derivative, \textit{i.e.},
\bea
\delta_{\hat \xi} V^M = {\mathcal L}_{\hat \xi} V^M =\hat \xi^N \partial_{N} V^M + (\partial^M \hat \xi_{P} - \partial_P \hat \xi^M) V^{P} + \omega \partial_{N} \hat \xi^N V^{M}\, .
\eea
In the previous expression we consider a generic parameter $\hat \xi^{M}$ and a density weight factor $\omega$. The generalized Lie derivative differs from the ordinary one since we need to ensure ${\cal L}_{\hat \xi} \eta_{M N}=0$. The closure of the generalized diffeomorphism transformations on an arbitrary generalized tensor $V^{M}{}_{N}$,
\bea
\Big[\delta_{\hat \xi_1},\delta_{\hat \xi_2} \Big] V^{M}{}_{N} = \delta_{\hat \xi_{21}} V^{M}{}_{N} \,,
\eea
is provided by the C-bracket,
\bea
\hat \xi^{M}_{12} = \hat \xi^{P}_{1} \frac{\partial \hat \xi^{M}_{2}}{\partial X^{P}} - \frac12 \hat \xi^{P}_{1} \frac{\partial \hat \xi_{2P}}{\partial X_{M}} - (1 \leftrightarrow 2) \, .
\label{Cbra}
\eea
An important comment here is that the closure of the generalized diffeomorphisms requires the strong constraint (\ref{SC}). Moreover, and from an algebraic point of view, DFT contains a non-trivial Jacobiator and therefore only satisfies the Jacobi identity `up to homotopy' \cite{Linf}.

On the other hand, the partial derivative of a generic double vector does not transform as a tensor. Therefore we define a covariant derivative as follows
\bea
D_{M} V^{N} = \partial_{M} V^{N} - \Gamma_{M}{}^{N}{}_{P} V^{P} \, .
\label{DFTcovder}
\eea
The compatibility $D_{M}\eta_{NP}=0$ implies $\Gamma_{MNP}=-\Gamma_{MPN}$. It is usual to impose \bea
{\cal T}_{MNP}=3 \Gamma_{[MNP]}=0 \, ,
\eea
since ${\cal T}_{MNP}$ plays the role of generalized torsion. A curious aspect of the double geometry is that there are not enough compatibility conditions to  fully determine $\Gamma_{MNP}$ \cite{RiemannDFT}, unlike in general relativity. Consequently, the generalized Riemann tensor cannot be fully determined but, nevertheless, a generalized Ricci tensor ${\cal R}_{MN}$ can be constructed as well as a generalized Ricci scalar ${\cal R}$, and these are fully determined in terms of the fundamental fields of DFT.

Another symmetry of DFT is the double Lorentz transformation that acts in the following way,
\bea
\delta_{\Lambda} V^{A} = V^{B} \Lambda_{B}{}^{A} \, ,
\eea
on an arbitrary flat vector $V^A$. Demanding $\delta_{\Lambda} \eta_{AB}=0$ we have $\Lambda_{AB}=-\Lambda_{BA}$. Moreover, using the decomposition/notation (\ref{projflat}), the condition $\delta_{\Lambda}{\cal H}_{AB}=0$ implies
\bea
\Lambda_{\un A \ov B} = \Lambda_{\ov A \un B} = 0 \, .
\eea
The partial derivative of an arbitrary flat vector $V^A$ does not transform as a tensor and therefore we introduce a flat covariant derivative,
\bea
D_{M} V^{A} = \partial_{M} V^{A} - W_{M}{}^{A}{}_{B} V^{B} \, ,
\eea
where $W_{MAB}=-W_{MBA}$  is the generalized spin connection. Imposing compatibility with the generalized frame we find that $W_{MAB}$ is not fully determined.

\item{\bf Fundamental fields:} The fundamental fields of DFT are a generalized frame $E_{MA}(X)=E_{MA}$ and a generalized dilaton $d(X)=d$. The former is equivalent to a vielbein for this double geometry and satisfies 
\bea
E_{M A} {\cal H}^{A B} E_{N B} & = & {\cal H}_{M N} \, , \\
E_{M A} \eta^{A B} E_{N B} & = & \eta_{M N} \, .
\eea
DFT can be written in terms of $d$ and ${\cal H}_{M N}$, the latter known as the generalized metric. In this case, the double Lorentz invariance is implicit. The generalized metric is an $O(D,D)$ element, \textit{i.e.}
\bea
{\cal H}_{M P} \eta^{P Q} {\cal H}_{Q N} = \eta_{M N} \, , 
\eea
and with the help of this dynamical metric and $\eta_{MN}$ one can define curved projectors,
\bea
P_{MN} = \frac{1}{2}\left(\eta_{MN} - {\cal  H}_{MN}\right) \, , \ \ \ \ \
\ov{P}_{MN} = \frac{1}{2}\left(\eta_{MN} + {\cal H}_{MN}\right)\ .
\label{proyc}
\eea
The previous projectors satisfy
\bea
&{\overline{P}}_{{M Q}} {\overline{P}}^{ Q}{}_{ N}={\overline{P}}_{{M N}}\, , &\quad {P}_{{M Q}} {P}^{Q}{}_{ N}={P}_{{M N}}, \nn\\
&{P}_{{M  Q}}{\overline{P}}^{Q}{}_{ N} = {\overline{P}}_{ {M Q}}  {P}^{ Q}{}_{ N} = 0\, ,  &\quad {\overline{P}}_{{MN}} + {P}_{{M N}} = \eta_{{M N}}\,,
\label{curveprojrel}
\eea
similar to the flat projectors. The main difference here is that $P_{AB}$ and $\ov P_{AB}$ are constants, while the curved ones are not for an arbitrary double background.

The fundamental fields transform with respect to generalized diffeomorphisms and double Lorentz transformations as follows,
\bea
\delta_{\hat \xi,\Lambda} E_{M A} & = & {\cal L}_{\hat \xi} E_{M A} + E_{M B} \Lambda^{B}{}_{A} \, ,  \\
\delta_{\hat \xi} d & = & \hat \xi^{N} \partial_N d - \frac12 \partial_{M} \hat \xi^{M} \, .
\eea
The generalized dilaton is a double Lorentz invariant, and its transformation under generalized diffeomorphisms is not covariant. However $e^{-2d}$ transforms as a generalized scalar density with $\omega=1$. On the other hand, both $E_{MA}$ and ${\cal H}_{MN}$ satisfy compatibility conditions 
\bea
D_{M}{\cal H}_{N P} & = & 0 \, , \quad
D_{M}{E}_{N P} = 0 \, .
\eea
The generalized fluxes are defined as 
\bea
F_{A B C} = 3 E_{[A} E^{M}{}_{B} E_{M C]} \, ,
\eea
with $E_{A}=\sqrt{2} E^M{}_{A} \partial_{M}$. It is possible to determine the totally antisymmetric part of the spin connection, $W_{ABC}\equiv \sqrt{2} W_{MBC} E^{M}{}_{A}$, in the following way
\bea
F_{A B C} = - 3 W_{[A B C]} \, .
\label{Fluxspin}
\eea
Finally, it is convenient to define 
\bea
F_{\un M \ov A \ov B} & = & \frac{1}{\sqrt{2}} E_{M}{}^{\un C} F_{\un C \ov A \ov B} \, , \\
F_{\ov M \un A \un B} & = & \frac{1}{\sqrt{2}} E_{M}{}^{\ov C} F_{\ov C \un A \un B} \, ,
\eea
for later use. The transformation rule of the previous objects is
\bea
\label{FluxM}
\delta_{\hat \xi, \Lambda} F_{\un M \ov A \ov B} & = & {\cal L}_{\hat \xi} F_{\un M \ov A \ov B} + \partial_{\un M} \Lambda_{\ov A \ov B} + 2 F_{\un M \ov C [\ov B} \Lambda^{\ov C}{}_{\ov A]} \, , \\
\delta_{\hat \xi, \Lambda} F_{\ov M \un A \un B} & = & {\cal L}_{\hat \xi} F_{\ov M \un A \un B} + \partial_{\ov M} \Lambda_{\un A \un B} + 2 F_{\ov M \un C [\un B} \Lambda^{\un C}{}_{\un A]} \, .
\eea
\item {\bf Action principle:} The action of DFT is given by
\bea
\int d^{2D}X e^{-2d} {\cal R}(E,d) \, ,
\eea
where ${\cal R}(E,d)$ is a two derivative scalar under generalized diffeomorphisms and it is invariant under Lorentz transformations. This object is known as the generalized Ricci scalar, and can be written in terms of the generalized fluxes \cite{flux},
\bea
{\cal R}(d,E_{MA}) & = & 2E_{\un{A}}F^{\un{A}} + F_{\un{A}}F^{\un{A}} - \frac16{F}_{\un{ABC}}F^{\un{ABC}} - \frac12{F}_{\ov{A}\un{BC}}F^{\ov{A}\un{BC}} \, ,
\label{GR}
\eea
where $F_{A} = \sqrt{2} \partial^{M} E_{M A} - 2 E_{A} d \, .$ 
\end{enumerate}

The present formulation is enough to recover the action principle (\ref{bos0}) and its symmetry rules, as we will see. For this reason, many extensions of DFT have been studied. Frameworks related to the canonical formulation of DFT can be found in \cite{CanonicalDFT}, the inclusion of the R-R sector can be found in \cite{RRDFT}, while supersymmetric extensions were worked out in \cite{SDFT}. Now we are interested in obtaining extra non-Abelian terms from DFT. Therefore, in the following part, we generalize the construction presented here.

\subsubsection{Heterotic/gauged double field theory}
The duality group of the heterotic/gauged formulation of DFT \cite{DFThetero} is $O(D,D+n)$ with $n$ the dimension of the gauge group. Since this group differs from $O(D,D)$, the construction described in the previous part requires the following substantial modifications:
\begin{enumerate}
\item {\bf Double geometry:} We need to use a double geometry with coordinates $X^{\cal M}=(\tilde{x}_{\mu}, x^{\mu}, x^{i})$ with ${\cal M}=0,\dots,2D-1+n$ and $i=1,\dots,n$. On each point of the extended double space we consider an extended double tangent space, so we are able to define flat vectors $V^{\cal A}$, with ${\cal A}$ a flat index, ${\cal A}=0,\dots,2D-1+n$. 

\item {\bf Symmetries:} The generalized diffeomorphisms now contain an extra term that depends on generalized structure constants $f_{\cal MNP}$,
\bea
\delta_{\hat \xi} V^{\cal M} = {\mathcal L}_{\hat \xi} V^{\cal M} =\hat \xi^{\cal N} \partial_{\cal N} V^{\cal M} + (\partial^{\cal M} \hat \xi_{\cal P} - \partial_{\cal P} \hat \xi^{\cal M}) V^{\cal P} + f^{\cal M}{}_{\cal NP} \hat \xi^{\cal N} V^{\cal P} + \omega \partial_{\cal N} \hat \xi^{\cal N} V^{\cal M} \, ,
\label{gaugeLie}
\eea
where $V^{\cal M}$ is an arbitrary double vector. The structure constants are fully antisymmetric and satisfy a Jacobi rule \cite{GDFT},
\bea
f_{{\cal MNP}}=f_{[{\cal MNP}]}\, , \qquad f_{[\cal MN}{}^{\cal R}f_{{\cal P}] {\cal R}}{}^{\cal Q}=0\, . \label{consf}
\eea

The closure of the transformations is given by a deformed bracket  
\bea
 [\hat \xi_1, \hat \xi_{2} ]^{ \cal M}_{(C_{f})}=2\hat \xi^{ \cal P}_{[1}\partial_{\cal P}\hat \xi_{2]}^{\cal M}-\hat \xi_{[1}^{ \cal N}\partial^{\cal M}\hat \xi_{2]\cal N}+f_{{\cal PQ}}{}^{\cal M} \hat \xi_{1}^{\cal P} \hat \xi_2^{\cal Q}\, ,
\eea
which reduces to the C-bracket when the structure constants vanish. Interesting enough, the closure requires the strong constraint (\ref{SC}) plus an extra constraint,
\bea
f_{\cal M N P} \partial^{\cal M} \star = 0 \, .
\label{Strongf}
\eea
We solve the new constraint considering that the generalized structure constants are non-vanishing only when ${\cal M,N,P}=i,j,k$, \textit{i.e.} $f_{\cal MNP}=f_{ijk}$, and $\partial_{i}=0$. 

The covariant derivative (\ref{DFTcovder}) remains unchanged, but $\Gamma_{\cal MNP}$ receives an extra term in its transformation rule
\bea
\delta_{\hat \xi} \Gamma_{\cal MNP} = \delta^{\rm ungauged}_{\hat \xi} \Gamma_{\cal MNP} + f_{\cal N Q P} \partial_{\cal M} \hat \xi^{\cal Q} \, ,
\eea
where the explicit form of the first term corresponds to an exercise (see Problem Sheet \ref{Ex2}). On the other hand, the double Lorentz symmetry remains unchanged.

\item{\bf Fundamental fields:} The fundamental fields of heterotic/gauged DFT are a generalized frame $E_{\cal M}{}^{\cal A}(X)$ and a generalized dilaton $d(X)$. The only modification at this point is related to the construction of the generalized fluxes $F_{\cal A B C}$, that now contain an extra term
\bea
F_{\cal A B C} = 3 E_{[\cal A} E^{\cal M}{}_{\cal B} E_{\cal M C]} + \sqrt{2} f_{\cal M N P} E^{\cal M}{}_{\cal A} E^{\cal N}{}_{\cal B} E^{\cal P}{}_{\cal C} \, .
\label{Gflux}
\eea
Similarly to the heterotic supergravity case we assume that $f_{\cal MNP}$ has the same units as derivatives.

\item {\bf Action principle:} The action of heterotic/gauged DFT is given by
\bea
\int d^{2D+n}X e^{-2d} {\cal R}(E,d) \, ,
\eea
where ${\cal R}(E,d)$ is a two derivative scalar under (\ref{gaugeLie}) and it is invariant under Lorentz transformations. This action principle now contains $f_{\cal MNP}$ contributions provided by the generalized fluxes (\ref{Gflux}) and, schematically, ${\cal R}$ has the same form as (\ref{GR}) but promoting the indices ${A}\rightarrow{\cal A}$.

\end{enumerate}

\subsubsection{Parametrization}
Here we present the parametrization of the different parameters, metrics and fields of heterotic/gauged DFT. The parametrization of the ungauged DFT is a particular case of this one. We follow the same structure of the previous part, step by step. 
\begin{enumerate}
\item {\bf Double geometry:} The parametrization of the coordinates is $X^{\cal M}=(\tilde{x}_{\mu}, x^{\mu}, x^{i})$. Since we solve the strong constraint using $\tilde \partial^{\mu} = \partial_{i}=0$, the parameters and fields of the resulting gauged supergravity do not depend on extra coordinates. The parametrization of the invariant metric is 
\be
{\eta}_{{\cal M N}}  = \left(\begin{matrix}\eta^{\mu\nu}&\eta^\mu{}_\nu&\eta^\mu{}_i\\ 
\eta_\mu{}^\nu&\eta_{\mu\nu}&\eta_{\mu i}\\\eta_{i}{}^\nu&\eta_{i\nu}&\eta_{ij}\end{matrix}\right)= \left(\begin{matrix}0&\delta^\mu{}_\nu&0\\ 
\delta_\mu{}^\nu&0&0\\0&0&\kappa_{ij}\end{matrix}\right) \ , \label{eta}
\ee
with $\mu, \nu=0,\dots, D-1$, $i,j=1,\dots, n$ and $\kappa_{ij}$ the Killing metric of the gauge group.  The flat indices can be split in the following way ${\cal A}=(\un a, \ov a, \ov i)$ where $\un a=1,\dots,D$, $\bar a=1,\dots,D$ and $\ov i$ runs from $1,\dots,n$. The latter is the flat version of the gauge index. The Lorentz invariant metrics are parametrized as 
\begin{equation}
\eta_{\cal A B} = \left(\begin{matrix} - \eta_{\un a \un b} & 0 & 0 \\ 0 & \eta_{\bar a \bar b} & 0 \\ 0 & 0 & \kappa_{\bar i \bar j} \end{matrix}\right) \ , \quad
{\cal H}_{\cal A B} = \left(\begin{matrix} \eta_{\un a \un b} & 0 & 0 \\ 0 & \eta_{\ov a \ov b} & 0 \\ 0 & 0 & \kappa_{\bar i \bar j} \end{matrix}\right) \ ,
\label{flatparam}
\end{equation}
where $\eta_{\un a \un b}$ and $\eta_{\ov a \ov b}$ can be identified with a flat and constant metric $\eta_{a b}$ 
\bea
\eta_{\bar a \bar b} \delta_{ab}^{\bar a \bar b} = \eta_{\un a \un b} \delta_{ab}^{\un a \un b} = \eta_{ab} \, , 
\eea
and $\kappa_{\bar i \bar j}$ is the flat version of the Killing metric. Defining the following constant object $e_{i}{}^{\bar i}$ we have a relation between the Cartan-Killing metric and its flat version,
\bea
\kappa_{i j} = e_{i}{}^{\bar i} \kappa_{\bar i \bar j} e_{j}{}^{\bar j} \, .
\eea
The same holds for the inverses of these metrics, defining $e^{i}{}_{\bar i}$ as the inverse of $e_{i}{}^{\bar i}$.

Before analyzing the symmetry rules, we include in this part the parametrization of the generalized frame,
\be
E^{\cal M}{}_{\cal A}  =\left(\begin{matrix}{ E}_{\mu \underline a}&  { E}^{\mu }{}_{\underline a} & E^i{}_{\underline a}\\ 
E_{\mu \overline  a}& E^\mu{}_{\overline  a}&E^i {}_{\overline a} \\
E_{\mu\overline i} &E^\mu{}_{\overline i} &E^i{}_{\overline i} \end{matrix}\right) \ = \
\frac{1}{\sqrt{2}}\left(\begin{matrix}-{ e}_{\mu \underline a}-C_{ \rho\mu} { e}^{\rho }{}_{\underline a} &  { e}^{\mu }{}_{\underline a} & -A_\rho{}^i { e}^{\rho }{}_{\underline{a}}\, , \\ 
e_{\mu \overline a}-C_{\rho \mu}{} e^{\rho }{}_{\overline{a}}& e^\mu{}_{\overline a}&-A_{\rho}{}^i  e^\rho{}_{\overline a} \\
\sqrt2 A_{\mu i}e^i{}_{\overline i} &0&\sqrt2 e^i{}_{\overline i} \end{matrix}\right)  \, ,
\label{Frame0}
\ee
where ${e}_{\mu \underline  a}$ and $e_{\mu \overline  a}$ are a pair of vielbeins for the same  metric tensor $g_{\mu \nu}$, i.e.,
\bea
e_{\mu}{}^{\ov a} \eta_{\ov a \ov b} e_{\nu}{}^{\ov b} & = & g_{\mu \nu} \, , \\
e_{\mu}{}^{\un a} \eta_{\un a \un b} e_{\nu}{}^{\un b} & = & g_{\mu \nu} \, .
\eea
We identify each vielbein (and their inverses) considering the following gauge fixing of the double Lorentz transformations to a single copy of Lorentz transformations,
\bea
{e}_{\mu \un a} \delta_{a}^{\un a} & = & {e}_{\mu \ov a} \delta_{a}^{\ov a} = e_{\mu a} \, , \\
{e}^{\mu}{}_{\un a} \delta_{a}^{\un a} & = & {e}^{\mu}{}_{\ov a} \delta_{a}^{\ov a} = e^{\mu}{}_{a} \, .
\eea
Finally in (\ref{Frame0}) we use the notation $C_{\mu \nu}=b_{\mu \nu} + \frac12 A_{\mu}{}^{i} A_{\nu i}$.

\item {\bf Symmetries:} The parametrization of the symmetry parameters is given by,
\bea
\hat \xi^{\cal M}& = & (\zeta_\mu, \xi^\mu, \lambda^i) \label{diffeosP}\\
\Lambda_{\ov a \ov b} & = & \Lambda_{a b} \delta_{\ov a \ov b}^{a b} \, , \quad 
\Lambda_{\un a \un b} = - \Lambda_{a b} \delta_{\un a \un b}^{a b} \\
\Lambda_{\ov a \ov{i}} & = & \label{gaugeuno} 0  \, , \quad
  \Lambda_{\ov{i} \ov{j}} =  f_{i j k} \lambda^{j} e^{k}{}_{\bar i} e^{i}{}_{\ov{j}}{} \, .
\eea
The components $\Lambda_{\ov a \ov{i}}$ and $\Lambda_{\ov i \ov{j}}$ are fixed to ensure $\delta E^{\mu}{}_{\ov i}=0$ and $\delta E^{i}{}_{\ov i}=0$ as required by (\ref{Frame0}). From (\ref{diffeosP}) we can understand how the generalized diffeomorphisms encode ordinary diffeomorphisms plus gauge transformations, while double Lorentz transformation encode the ordinary Lorentz transformations. For instance, from $\delta E^{\mu \ov a}$ we obtain 
\be
\delta E^{\mu}{}_{\ov a} \delta^{\ov a}_{a} = \delta e^{\mu}{}_{a} =  \xi^{\rho}\partial_{\rho}  e^{\mu}{}_{a}  - \partial_{\rho}\xi^{\mu} e^{\rho}{}_{a} + e^{\mu}{}_{b} \Lambda^{b}{}_{a} \, ,
\label{inverseparam}
\ee
in agreement with (\ref{inverse0}). 

\item{\bf Fundamental fields:} The parametrization of the generalized frame was given in (\ref{Frame0}), while the parametrization of the generalized metric, 
\bea
{\cal H}_{\cal M N} = \left(\begin{matrix} g^{\mu \nu} & - g^{\mu \rho} C_{\rho \nu} & - g^{\mu \rho} A_{\rho i} \\
- C_{\rho \mu} g^{\nu \rho}  & g_{\mu \nu} + C_{\rho \mu} C_{\sigma \nu} g^{\rho \sigma} + A_{\mu}{}^i \kappa_{ij} A_{\nu}{}^j &
C_{\rho \mu} g^{\rho \sigma} A_{\sigma i} + A_{\mu}{}^j \kappa_{ji} \\
- g^{\nu \rho} A_{\rho i} & C_{\rho \nu} g^{\rho \sigma} A_{\sigma i} +  A_{\nu}{}^j \kappa_{ij} & \kappa_{ij} + A_{\rho i} g^{\rho \sigma} A_{\sigma j}\end{matrix}\right) \ ,
\label{generalizedmetric}
\eea
can be easily obtained from the $E_{\cal M}{}^{\cal A} {\cal H}_{\cal A B} E_{\cal N}{}^{\cal B}={\cal H}_{\cal MN}$. This parametrization agrees with the results of \cite{Maharana}.  

The generalized dilaton is given by
\bea
e^{-2d} = \sqrt{-g} e^{-2\phi} \, .
\eea
Using the parametrization of the generalized frame and dilaton, it is straightforward to compute the parametrization of the different projections of the generalized fluxes \cite{LNR},
\bea \label{f1}
F_{\overline a\underline{ bc}} & = & -
w^{(+)}_{ a bc} \delta_{\overline a\underline{ bc}}^{abc}\, ,\\
F_{\underline  a \overline{bc}} & = & w^{(-)}_{ a bc} \delta_{\underline  a \overline{bc}}^{abc}\, ,\\
F_{\overline{a bc}} & = & 3\left(w_{[{abc}]}-\frac16 \bar H_{{abc}}\right) \delta_{\overline{a bc}}^{abc}
\, ,\\
F_{\underline {a bc}}& = & -3\left(w_{[abc]}+\frac16 \bar H_{abc}\right) \delta_{\underline {a bc}}^{abc}\, ,\\
F_{\overline i \underline{ab}} & = & - \frac1{\sqrt2} e^ \mu{}_a e^\nu{}_b e{}_{i\overline i} F^i_{\mu\nu} \delta_{\underline{ab}}^{ab}\, , \label{Fluxg}\\
F_{\underline a \overline i \overline j} & = & -e^i{}_{\overline i} e^j{}_{\overline j} e^\mu{}_a A_{\mu}{}^{k} f_{ijk} \delta_{\underline a}^{a} \, ,\\
F_{\overline{ijk}} & = & \sqrt2e^i{}_{\overline i }e^j{}_{\overline j}e^k{}_{\overline k }f_{ijk}\, ,\\
F_{\underline a} & = & \left(\partial_\mu e_a^\mu+e_a^\mu e_b^\nu\partial_\mu e^b_\nu-2e^\mu{}_{ a}\partial_\mu\phi\right) \delta_{\un a}^{a} \label{f2}\, ,
\eea
where 
\bea
\bar H_{abc}&=& 3e^\mu{}_{a} e^\nu{}_b e^\rho{}_{c} \left(\partial_{[\mu}b_{\nu\rho]}-A_{[\mu}^i\partial_\nu A_{\rho]i}+\frac13
f_{ijk}A_{\mu}{}^i A_\nu{}^j A_\rho{}^k\right)\, .\label{flatH}
\eea

\item {\bf Action principle:} The action of heterotic/gauged DFT reduces to (\ref{het0}) after parametrization. For example, the last term of the heterotic/gauged DFT action principle, $-\frac12 F_{\bar i \un b \un c} F^{\bar i \un b \un c}$, reproduces the term $-\frac14 F_{\mu \nu}{}^{i} F^{\mu \nu}{}_{i}$ when the component $F_{\ov i \un b \un c}$ is replaced by (\ref{Fluxg}).

\end{enumerate}

\subsection{Part B: the generalized Green-Schwarz mechanism}
\label{B2}
\subsubsection{Heterotic DFT}

In the first lecture we showed that the $b$-field transforms non-trivially under Lorentz transformations when four-derivative terms are considered in the heterotic supergravity action principle (we turn off the b-parameter),
\bea
\delta_{\Lambda} b_{\mu \nu} = \frac{a}{2} \partial_{[\mu} \Lambda^{ab} w^{(-)}_{\nu]ab} \, ,
\eea
where in the last expression we chose the Bergshoeef-de Roo approach (\ref{Lorentz1}) keeping $a$ undetermined. Our goal now is to deform the transformation of the generalized frame in order to construct a higher-derivative DFT formulation mimicking this behavior \footnote{We turn off the gauged part, for simplicity. Once the DFT is deformed, we can promote $O(D,D)\rightarrow O(D,D+n)$.}. The idea is to consider a generalized Green-Schwarz mechanism at the DFT level,
\bea
\delta_{\Lambda} E_{M}{}^{\ov A} & = & E_{M}{}^{\ov B} \Lambda_{\ov B}{}^{\ov A} + \frac{a}{2} T_{1 M}{}^{\ov A} \, , \\  \delta_{\Lambda} E_{M}{}^{\un A} & = & E_{M}{}^{\un B} \Lambda_{\un B}{}^{\un A} + \frac{a}{2} T_{2 M}{}^{\un A} \, ,
\eea
where the natural proposal is 
\bea
T_{1 M}{}^{\ov A} = \partial_{P} \Lambda^{\ov B \ov C} F_{\un M \ov B \ov C} E^{P \ov A} \, ,
\eea
since  $F_{\un M \ov B \ov C}$ is related to $w^{(-)}_{\mu b c}$. Before testing this ansatz, let us show that the curved index of a generic first-order correction to the generalized frame transformation must satisfy
\bea
\label{projcond1}
T_{1 M \ov A} & = & T_{1 \un M \ov A} \, , \\
T_{2 M \un A} & = & T_{2 \ov M \un A} \, .
\label{projcond2}
\eea
Any first order-transformation $\delta^{(1)}_{\Lambda}$ has to ensure the following compatibility condition,
\bea
\delta^{(1)}_{\Lambda}(E_{M}{}^{\ov A} E^{M \un B}) & = & 0 \, ,
\eea
and then (\ref{projcond1}) and (\ref{projcond2}) are mandatory. Moreover, both relations are compatible with 
\bea
\delta^{(1)}_{\Lambda} \eta_{A B} & = & 0 \, , \\
\delta^{(1)}_{\Lambda} {\cal H}_{A B} & = & 0 \, .
\eea
Finally, from $\delta^{(1)}_{\Lambda} \eta_{M N}$ we have
\bea
\delta^{(1)}_{\Lambda}(E_{M}{}^{\ov A} E_{N \ov A} + E_{M}{}^{\un A} E_{N \un A} ) = 0  \, ,
\eea
and, consequently, $T_{2 \ov M \un A}$ must be constructed from $T_{1 \un M \ov A}$ according to
\bea
T_{2(M}{}^{\un A} E_{N)\un A} = - T_{1(M}{}^{\ov A} E_{N) \ov A} \, .
\eea
Then, 
\bea
T_{2 M}{}^{\un A}= - \partial_{\ov M} \Lambda^{\ov A \ov B} F_{\un N \ov A \ov B} E^{N \un A} \, ,
\eea
and the generalized Green-Schwarz mechanism can be written as \footnote{Convention: $[\ov N \, \un M] = \frac12 \ov N \, \un M - \frac12 \ov M \, \un N$ (we exchange indices but projections remain unchanged).} \cite{Tduality} \cite{GreenDFT}
\bea
\delta_{\Lambda} E_{M}{}^{A} = E_{M}{}^{B} \Lambda_{B}{}^{A} + a \partial_{[\ov N} \Lambda^{\ov B \ov C} F_{\un M] \ov B \ov C} E^{N A} \, .
\eea
On the other hand the generalized dilaton does not have a first order transformation. Now we can test the ansatz in order to see if the Green-Schwarz mechanism for the $b$-field is inherited. For simplicity we will study ungauged DFT, so the duality group is $O(D,D)$. However, a crucial question is: how do we know that the fields that we use to parametrize the generalized frame in (\ref{Frame0}), $\Big\{e,b\Big\}$, are the same fields as the ones in the Bergshoeff-de Roo approach, $\Big\{e_{\rm BdR},b_{\rm BdR}\Big\}$? We have seen, for example, that the previous set differs from that used in the Metsaev-Tseytlin approach, $\Big\{e_{\rm MT},b_{\rm MT} \Big\}$ , by a $b$-field redefinition, as we showed in (\ref{bequivalence}). Consequently, we do not know if our set of fields agrees with other approaches. Moreover, the set of fields that we are using to parametrize the DFT fields are covariant under Buscher transformations \cite{Buscher}, so it is very likely that field redefinitions will be required. 

The safest way to go ahead is to use an arbitrary set of fields, $\Big\{\tilde e,\tilde b\Big\}$, and to consider a new parametrization for the generalized frame,    
\be
E^{M}{}_{A}  = 
\frac{1}{\sqrt{2}}\left(\begin{matrix}-{ \tilde e}_{\mu \underline a}- \tilde b_{ \rho\mu} {\tilde e}^{\rho }{}_{\underline a} &  {\tilde e}^{\mu }{}_{\underline a} \, , \\ 
\tilde e_{\mu \overline a}- \tilde b_{\rho \mu}{} \tilde e^{\rho }{}_{\overline{a}}& \tilde e^\mu{}_{\overline a}
\end{matrix}\right)  \, .
\label{Frame1}
\ee
If we want to obtain the transformation rule $\delta \tilde e^{\mu \bar a}$, we need to inspect $\delta E^{\mu \bar a}$, 
\bea
\delta E^{\mu \bar a} = E^{\mu \bar b} \lambda_{\bar b}{}^{\bar a} + \frac{a}{2} \bar P_{P}{}^{Q} \partial_{Q} \Lambda^{\ov b \ov c} (\frac{1}{\sqrt{2}} E_{M}{}^{\un a} F_{\un a \ov b \ov c}) E^{P \ov a} \, .
\eea
Parametrizing the previous expression we get,
\bea
\delta \tilde e^{\mu a} = \tilde e^{\mu b} \Lambda_{b}{}^{a} - \frac{a}{4} \tilde \partial^{a} \Lambda^{bc} \tilde w^{(-)}_{dbc} \tilde e^{\mu d} \, ,
\eea
where $\tilde \partial_{a} = \tilde e^{\mu}{}_{a} \partial_{\mu}$ and $\tilde w_{abc}=w_{abc}(\tilde e)$. Let us observe that due to the generalized Green-Schwarz mechanism, the inverse vielbein $\tilde e^{\mu}{}_{a}$ obtains an undesired transformation as well as $\tilde e_{\mu a}$. The latter can be easily computed considering
\bea
\delta_{\Lambda}(\tilde e^{\mu a} \tilde e_{\mu}{}^{b}) = \delta^{(1)}_{\Lambda}(\eta^{a b}) = 0 \, ,
\eea
and we get
\bea
\delta \tilde e_{\mu a} = \tilde e_{\mu b} \Lambda^{b}{}_{a} + \frac{a}{4} \partial_{\mu} \Lambda^{bc} \tilde w^{(-)}_{abc} \, .
\eea
Since we want to match with the set of fields used by Bergshoeff-de Roo we need to propose the following field redefinition for the vielbein,
\bea
e_{\rm BdR}^{\mu a} & = & \tilde e^{\mu a} + \frac{a}{8} \tilde e^{\mu b} \tilde w^{(-)}_{b}{}^{cd} \tilde w^{(-)}_{\nu cd} \tilde e^{\nu a} \, , \\
e^{\rm BdR}_{\mu a} & = & \tilde e_{\mu a} - \frac{a}{8} \tilde w^{(-)}_{a}{}^{cd} \tilde w^{(-)}_{\mu cd} \, ,
\label{vielbeinredef}
\eea
and similarly for the Lorentz parameter,
\bea
\Lambda^{\rm BdR}_{ab} = \Lambda_{ab} - \frac{a}{4} \tilde w^{(-)}_{[a}{}^{cd} \tilde \partial_{b]} \Lambda_{cd} \, .  
\eea
Inspecting the transformation rule of the $E_{\mu \bar a}$ it is straightforward to obtain
\bea
\delta \tilde b_{\mu \nu} = \frac{a}{2} \partial_{[\mu} \Lambda^{ab} \tilde w^{(-)}_{\nu]ab}
\label{GSbDFT2}
\eea
which is exactly the transformation of the Bergshoeff-de Roo approach presented in the previous lecture\footnote{The way to match with the fields presented in the original publication \cite{BdR} is very well discussed in the appendix of \cite{LNR}.} when we turn off the b-parameter. Some readers may be worried about the $\tilde{w}_{\nu ab}$ in (\ref{GSbDFT2}) but the difference between ${w}_{\nu ab}$ and $\tilde{w}_{\nu ab}$ can be neglected at this order since it introduces $\alpha'^2$ contributions to (\ref{GSbDFT2}). Then $\tilde b_{\mu \nu}=b^{\rm BdR}_{\mu \nu}$ and no more field redefinitions are required. This generalized Green-Schwarz mechanism agrees with \cite{Heteroalpha}. In the next part we will generalize this method to include the bi-parametric case.

\subsubsection{Bi-parametric extension}
In the previous part we have given some arguments to propose a generalized Green-Schwarz transformation at the heterotic DFT level. We mainly follow \cite{Tduality}, where the following bi-parametric proposal was analyzed in detail,
\bea
\delta_{\Lambda} E_{M}{}^{A} = E_{M}{}^{B} \Lambda_{B}{}^{A} + a \partial_{[\ov P} \Lambda^{\ov B \ov C} F_{\un M] \ov B \ov C} E^{P A} + b \partial_{[\un P} \Lambda^{\un B \un C} F_{\ov M] \un B \un C} E^{P A} \, .
\eea
Here the parameters are related to (super)gravity formulations when
 \bea
 \label{casesdft}
 (a,b) = 
   \begin{cases} 
      (-1,-1)  & \mbox{bosonic DFT \, , }  \\
      (-1,0)  & \mbox{heterotic DFT \, ,} \\ 
      (0,0) & \mbox{type II DFT \, .} 
   \end{cases}
\eea
It is important to clarify that the previous parameters allow us to access the ungauged bosonic supergravity Lagrangians and their gravitational corrrections. A higher-derivative study related to $Z_{2}$-transformations in DFT was done in \cite{Z2DFT}, preceding \cite{Tduality}. In \cite{Z2DFT} the combination (-1,-1) was referred to as DFT$^{(+)}$ and the combination (-1,1) as DFT$^{(-)}$. We leave some comments related to this theory for the next part \footnote{Nowadays some papers use the name HSZ for Hohm, Siegel and Zwiebach after the authors of \cite{HSZ0} for the combination (-1,1), which is a very particular higher-derivative extension of DFT.}.

Let's see what field redefinitions are required in the bi-parametric case. Parametrizing $\delta_{\Lambda} E^{\mu a}$ we have,
\bea
\delta \tilde e^{\mu a} & = & \tilde e^{\mu b} \Lambda_{b}{}^{a} - \frac{a}{4} \tilde \partial^{a} \Lambda^{bc} \tilde w^{(-)}_{dbc} \tilde e^{\mu d} + \frac{b}{4} \tilde \partial^{\mu} \Lambda^{b c} \tilde w^{(+)}_{\nu b c} \tilde e^{\nu a} \, .
\eea
Therefore, the field redefinition for the vielbein now is \footnote{Once the vielbein redefinition is imposed, then we must propose a dilaton redefinition according to $e^{-2d}=\sqrt{- \tilde g} e^{-2 \tilde \phi} = \sqrt{-g} e^{-2 \phi} $.},
\bea
e_{\rm BdR}^{\mu a} & = & \tilde e^{\mu a} + \frac{a}{8} \tilde e^{\mu b} \tilde w^{(-)}_{b}{}^{cd} \tilde w^{(-)}_{\nu cd} \tilde e^{\nu a} - \frac{b}{8} \tilde e^{\mu b} \tilde w^{(+)}_{b}{}^{cd} \tilde w^{(+)}_{\nu cd} \tilde e^{\nu a} \, , \\ 
e^{\rm BdR}_{\mu a} & = & \tilde e_{\mu a} - \frac{a}{8} \tilde w^{(-)}_{a}{}^{cd} \tilde w^{(-)}_{\mu cd} + \frac{b}{8} \tilde w^{(+)}_{a}{}^{cd} \tilde w^{(+)}_{\mu cd} \, ,
\eea
while the Lorentz parameter field also needs a redefinition,
\bea
\Lambda^{\rm BdR}_{ab} = \Lambda_{ab} - \frac{a}{4} \tilde w^{(-)}_{[a}{}^{cd} \tilde \partial_{b]} \Lambda_{cd} + \frac{b}{4} \tilde w^{(+)}_{[a}{}^{cd} \tilde \partial_{b]} \Lambda_{cd}.  
\eea
Finally the $b$-field transformation is
\bea
\delta \tilde b_{\mu \nu} = \frac{a}{2} \partial_{[\mu} \Lambda^{ab} \tilde w^{(-)}_{\nu]ab} - \frac{b}{2} \partial_{[\mu} \Lambda^{ab} \tilde w^{(+)}_{\nu]ab} \, .
\label{GSbDFT}
\eea
Up to this point we have deformed the double Lorentz transformations in a consistent way. Moreover, it is straightforward to verify that the two-parameter deformation closes as follows
\bea
[\delta_{\Lambda_{1}},\delta_{\Lambda_{2}}] E_{M A} = \delta_{\Lambda_{21},\xi_{21}} E_{M A} \, .
\label{highclosure}
\eea
The previous computation is part of the Problem Sheet 2.3. However we still need a recipe to construct the four-derivative action at the DFT level,
\bea
S_{DFT} = \int d^{2D}X e^{-2d} ({\cal R} + a {\cal R}^{(-)} + b {\cal R}^{(+)}) \, .
\label{actionfull}
\eea
Since $\delta^{(1)}_{\Lambda}{\cal R}\neq0$, then ${\cal R}^{(\pm)}\neq0$ to ensure the invariance of the action. We leave this point to discuss in the third lecture.

\subsubsection{Comments about HSZ theory}
All $(a,b)$ combinations different from $(-1,-1)$, $(-1,0)$,$(0,0)$ produce $O(D,D)$ invariant theories that are not string theories. One example is HSZ theory, where $(a,b)=(-1,1)$, originally constructed in \cite{HSZ0} and explored in \cite{HSZ} \cite{Usman} \cite{DiegoyEric}. This theory can  be written in the generalized metric formulation, where Lorentz invariance is implicit. In this case the generalized diffeomorphims are deformed as,
\bea
\delta_{\hat \xi} {\cal H}_{M N} = && {\cal L}_{\hat \xi} {\cal H}_{M N} - \frac12 \partial_{(M}{\cal H}^{PQ} \partial_{P}K_{Q|N)} + \frac12 {\cal H}_{(M}{}^{P} {\cal H}_{N)}{}^{Q} \partial_{P}{\cal H}^{RS} \partial_{R}K_{SQ} \nn \\ 
&& - \partial_{P} {\cal H}_{Q(M}\partial_{N)}K^{QP} + {\cal H}_{( M}{}^{P} {\cal H}_{N)}{}^{Q} \partial_{R} {\cal H}_{SP} \partial_{Q} K^{SR} \, ,
\label{HSZdiff}
\eea
where $K_{MN}=2\partial_{[M} \hat \xi_{N]}$. On the other hand, the generalized dilaton does not have a first order transformation.

In HSZ the C-bracket is consistently deformed,
\begin{equation}
\left[\delta_{\hat \xi_1}\, , \, \delta_{\hat \xi_2}\right] = \delta_{[\hat \xi_2 ,\hat \xi_1]_{C'}} \ ,
\end{equation}
where
\begin{equation}
\left[\hat \xi_1 , \hat \xi_2\right]_{C'}^M = 2\hat \xi^{ P}_{[1}\partial_{ P}\hat \xi_{2]}^{ M}-\hat \xi_{[1}^{ N}\partial^{ M}\hat \xi_{2] N} + \partial_P \hat \xi_{[1}^Q \partial^M \partial_Q \hat \xi_{2]}^P \ .
\label{Chsz}
\end{equation}
The four-derivative action principle of the HSZ theory at the supergravity level is given by,
\bea
S = \int d^Dx \sqrt{-g} e^{-2 \phi}\left(R  + 4 \partial_\mu \phi \partial^\mu \phi - \frac 1 {12} \widehat H_{\mu \nu \rho} \widehat H^{\mu \nu \rho} \right) \ , \label{ActionHSZ}
\eea
where
 \be
\widehat H_{\mu \nu \rho} = 3\, \partial_{[\mu} b_{\nu \rho]} + 3 \Omega_{\mu \nu \rho} \ , \ \ \ \ \Omega_{\mu \nu \rho} = \Gamma_{[\mu| \sigma|}^\delta \partial_{\nu} \Gamma_{\rho] \delta}^\sigma + \frac 2 3
\Gamma_{[\mu | \sigma|}^\delta \Gamma_{\nu| \lambda|}^\sigma \Gamma_{\rho] \delta}^\lambda \ . \label{Hhat}
\ee
In this formulation, the Chern-Simons terms deform the 3-form through Christoffel connections and the $b$-field transformation is corrected according to
\be
\delta^{(1)}_{\xi} b_{\mu \nu} = \partial_{[\mu} \partial_{|\rho|} \xi^\sigma \Gamma_{\nu]\sigma}^\rho \ . \label{GSHSZ}
\ee
The Green-Schwarz mechanism appears as a correction to the diffeomorphism transformation for this field. Both the dilaton and the metric tensor do not receive corrections to their transformations. 

While HSZ theory does not correspond to a string theory, it could be an important theory to understand the $\alpha'$ structure of string theory for several reasons:
\begin{enumerate}
    \item The full higher-derivative Lagrangian (as a deformed DFT) is known to all orders\footnote{Strictly speaking, we cannot say that this is an $\alpha'$ expansion, since this is not a string theory.}. It was computed in \cite{DiegoyEric} in terms of a double metric ${\cal M}_{MN}$ (which is not an $O(D,D)$ element) and a generalized dilaton $d$. It contains terms up to six derivatives. 
    
\item The odd higher-derivative terms of HSZ theory (in powers of $\alpha'$, i.e., 4-derivative terms, 8-derivative terms, and so on), at the supergravity level, have the same structure of the $\alpha'$-corrections with odd $b$-field for heterotic supergravity. For this reason HSZ does not have $(\rm Riem)^2$ terms, but it has Chern-Simons contributions.

\item The even higher-derivative terms of this theory (in powers of $\alpha'$, i.e., 6-derivative terms, 10-derivative terms, and so on), at the supergravity level, have the same structure of the $\alpha'$-corrections with even $b$-field for bosonic supergravity (up to an overall sign). For this reason HSZ has $(\rm Riem)^3$ terms. The following action reproduces the on-shell cubic amplitudes of HSZ theory \cite{Usman},
\bea
S &=& \int d^Dx \sqrt{-g} e^{-2 \phi}(R + 4 \nabla_\mu \phi \nabla^\mu \phi - \frac 1 {12} \widehat H_{\mu \nu \rho} \widehat H^{\mu \nu \rho} \nn \\ && - \frac 1 {48}\, R_{\mu \nu}{}^{\alpha \beta} R_{\alpha \beta}{}^{\rho \sigma} R_{\rho \sigma}{}^{\mu \nu}) \, .  \label{Action2ndOrderNZ}
\eea
The coefficient of the Riemann-cubed term is minus the coefficient of the same term in bosonic string theory and no such term appears in the heterotic string.  It was shown in \cite{Camanho:2014apa} that interactions that contribute to three-point amplitudes with three external gravitons lead to causality violations which require the existence of an infinite tower of new particles with spin higher than two. Such violations are avoided when the full structure of string theory is taken into account \cite{D'Appollonio:2015gpa}. In bosonic string theory there is also a non-zero Gauss-Bonnet Riemann-cubed term $G_3$
\be
G_3 = R_{\mu \nu}{}^{\alpha \beta} R_{\alpha \beta}{}^{\rho \sigma} R_{\rho \sigma}{}^{\mu \nu} - 2\, R^{\mu \nu \alpha \beta} R_{\nu \lambda \beta \gamma} R^\lambda{}_\mu{}^\gamma{}_\alpha \ ,
\ee
but its presence can only be seen from four-point amplitudes \cite{Metsaev:1986yb}. These terms (up to a sign) are part of the HSZ Lagrangian, as shown in \cite{DiegoyEric}.

\end{enumerate}

\newpage

\subsection{Exercises}
\label{Ex2}
\begin{enumerate}
\item Consider a generic double vector $V^{M}$ and show that $\Big[\delta_{\hat \xi_1},\delta_{\hat \xi_2} \Big] V^{M} = \delta_{\hat \xi_{21}} V^{M}$ holds (consider both ungauged/gauged DFT).

    \item Decide if the following projections are determined or undetermined: $\Gamma_{\un M \un N \un P}$, $\Gamma_{\un M \un N \ov P}$, $\Gamma_{\un M \ov N \ov P}$, $\Gamma_{\ov M \ov N \ov P }$, and write the determined projections in terms of ${\cal H}_{MN}$.

\item Demanding $
\int d^{2D}X e^{-2d} V \nabla_{M} V^{M}  = - \int d^{2D}X e^{-2d} V^{M} \nabla_{M} V $ show that the trace of the connection is $\Gamma_{M N}{}^{M} = -2\partial_{N} d$. Then show that $W_{A B}{}^{A}=-F_{B}$.

\item Find $\delta_{\hat \xi} \Gamma_{MNP}$. Then show that $\Gamma_{[MNP]}$
transforms as a generalized tensor. 

\item Show that (\ref{f1})-(\ref{f2}) hold.

\item Using the parametrization (\ref{Frame0}), show that $\delta E^{\mu}{}_{\bar i}=\delta E^{i}{}_{\bar i}=0$ implies (\ref{gaugeuno}).

\item Show that $\delta E^{\mu}{}_{\un a} \delta_{a}^{\un a}$ also reproduces (\ref{inverseparam}).

\item Compute the transformation rule for $b_{\mu \nu}$ and $A_{\mu i}$ coming from heterotic/gauged DFT.

\item * Show that the ungauged DFT action reduces to (\ref{bos0}) after parametrization.

\item Show that the redefinitions (\ref{vielbeinredef}) trivialize the transformation of the Bergshoeff-de Roo vielbein up to a Lorentz transformation.

\item Compute $\delta \tilde b_{\mu \nu}$ from $\delta E_{\mu a}$ in order to obtain (\ref{GSbDFT}).

\item Compute $\Lambda_{21}$ and $\hat \xi_{21}$ to show that (\ref{highclosure}) holds. 

\item * Show that $\delta_{\Lambda}{\cal R}\neq0$ when we deformed the generalized frame double Lorentz transformation. Impose $b=0$ (heterotic DFT) to simplify the computation, and then generalize the result to the bi-parametric case.

\item Compute the closure of the deformed generalized diffeomorphisms acting on ${\cal H}_{MN}$ in HSZ theory.

\item Show that $\delta_{\hat \xi}({\cal H}_{M P} \eta^{P Q} {\cal H}_{Q N})=0$ in HSZ theory.

\end{enumerate}
\newpage

\section{Lecture 3: Extended space and higher-order formulations}
\label{Sec3}
\subsection{Part A: a systematic procedure to first-order formulations}
\label{A3}
In the previous lecture we showed the form of a suitable generalization of the Green-Schwarz mechanism at the DFT level. Our starting point was an ansatz for the deformations of the double Lorentz transformations acting on the generalized frame, and we proceeded very carefully to satisfy all the DFT constraints. These transformations imply new four-derivative terms in the DFT Lagrangian, but we still do not have a method to compute them. For this reason, now we study a systematic procedure to reproduce both the symmetries and the higher-derivative corrections of the DFT Lagrangian. This procedure was originally presented in \cite{gbdr} together with a  generalization to higher-order corrections in heterotic DFT. We will discuss this point and the state of the art of this topic in the last part.

\subsubsection{Proposal}
As we have seen in (\ref{B1}), it is possible to construct gravitational $\alpha'$-corrections considering a map between gauge and Lorentz symmetries, provided by gauge generators. This relation was first observed by Bergshoeff and de Roo \cite{BdR}, and the assignments were $A_{\mu i}\rightarrow \frac{1}{\sqrt 2}w^{(-)}_{\mu a b}$ and $\lambda^{i}\rightarrow - \frac{1}{\sqrt 2} \Lambda^{ab}$. Now we want to generalized these identifications at the DFT level, but we deal with the following problem: the $O(D,D+K)$ formulation of gauged DFT does not have a generalized gauge field or parameter to perform the identifications. Here we focus on an arbitrary gauge group of dimension $K$. 

A possible solution to the previous problem is to break the $O(D,D+K)$ representations in terms of $O(D,D)$ representations \cite{HSZhetero}. This procedure allows us to obtain a generalized gauged field/parameter at the DFT level. We start by considering a generalized frame for gauged DFT ${\cal E}_{\cal M }{}^{\cal A}$, where ${\cal M}= 1,\dots,2D-1+K$ and ${\cal A}= 1,\dots, 2D+K$. Keeping the notation of Lecture 2, we can split the extended indices as ${\cal M}=(M,\alpha)$ and ${\cal A}=(A,\ov \alpha)$, where $\alpha$ and $\ov \alpha$ are curved/flat gauge indices that run from $1,\dots,K\, ,$ respectively. The $O(D,D+K)$ frame can be parametrized in terms of $O(D,D)$ multiplets as
\be
{\cal E}_{\cal M}{}^{\cal A}  = 
\left(\begin{matrix}{\cal E}_{M}{}^{A}& {\cal E}_{M}{}^{\bar \alpha}\\ 
{\cal E}_\alpha{}^A&{\cal E}_{\alpha}{}^{\bar \alpha} 
\end{matrix}\right)
= \left(\begin{matrix}({\chi}^{\frac12})_{M}{}^{N} E_{N}{}^{A}& -A_{M}{}^{\beta} e_{\beta}{}^{\ov \alpha}\\ 
{A}^{M}{}_\alpha E_{M}{}^{A}&({\Box}^{\frac12})_{\alpha}{}^{\beta} e_{\beta}{}^{\ov \alpha} 
\end{matrix}\right) \ , \label{Extframe}
\ee
where $e_{\alpha}{}^{\ov \a}$ (and its inverse $e^{\alpha}{}_{\ov \a}$) are constant objects that are related to the Cartan-Killing metric $\kappa_{\a \b}$ and its flat version $\kappa_{\ov \a \ov \b}$ as
\bea
e_{\alpha}{}^{\ov \a} \kappa_{\ov \a \ov \b} e_{\b}{}^{\ov \b} & = & \kappa_{\a \b} \, , \\
e^{\alpha}{}_{\ov \a} \kappa_{\a \b} e^{\b}{}_{\ov \b} & = & \kappa_{\ov \a \ov \b}
\, .
\eea
In (\ref{Extframe}) we defined the following objects,
\bea
\label{chiex}
{\chi}_{M N}=\eta_{M N} - A_{M}{}^{\alpha} A_{N \alpha} &\longrightarrow& {(\chi^{\frac12}})_{M N}=\eta_{M N} - \frac12 A_{M}{}^{\alpha} A_{N \alpha} + {\cal O}(A^4) \, , \\
{\Box}_{\a \b}=\kappa_{\a \b} - A_{M \alpha} A^{M}{}_{\beta} &\longrightarrow& {(\Box^{\frac12})}_{\a \b}=\kappa_{\a \b} - \frac12 A_{M \alpha} A^{M}{}_{\beta} + {\cal O}(A^4) \, ,
\eea
and we impose
\bea
E^{M \ov A} A_{M \a} = 0 \, ,
\eea
or equivalently
\bea
A_{M \a}=A_{\un M \a}, 
\eea
in order to eliminate extra degrees of freedom. From the previous condition we need ${\cal E}_{\alpha}{}^{\ov A}=0$, which will require a suitable gauge fixing. On the other hand, the projectors ${\cal P}_{\cal M N}$, $\overline{\cal P}_{\cal M N}$ and ${\cal P}_{\cal A B}$, $\overline{\cal P}_{\cal A B}$ can be decomposed as 
\be
{\cal P}_{\cal M N}  =\left(\begin{matrix}{P}_{M N}&  0 & \ \ 0 \\ 
0 & 0 & \ \ 0 \\
0 & 0 & \ \ 0 
\end{matrix}\right) \, , \quad \ov {\cal P}_{\cal M N}  =\left(\begin{matrix}0 &  0 & 0\\ 
0 & {\ov P}_{M N} & 0 \\
0 & 0 & {\kappa}_{\a \b} 
\end{matrix}\right) \, ,
\ee
\be
{\cal P}_{\cal A B}  =\left(\begin{matrix}{P}_{A B}&  0 & \ \ 0 \\ 
0 & 0 & \ \ 0 \\
0 & 0 & \ \ 0 
\end{matrix}\right) \, , \quad \ov {\cal P}_{\cal A B}  =\left(\begin{matrix}0 &  0 & 0\\ 
0 & {\ov P}_{A B} & 0 \\
0 & 0 & {\kappa}_{\ov \a \ov \b} 
\end{matrix}\right) \, .
\ee

From an $O(D,D)$ perspective, $\Big\{ E_{M A}, A_{M \alpha}, d \Big\}$ are now the fundamental fields of the gauged DFT. Similarly to the field decomposition (\ref{Extframe}), the generalized diffeomorphism transformations given by $\hat \xi_{\cal M}$, and the double Lorentz transformations given by $\Gamma_{\cal A \cal B}$, must be decomposed in a consistent way \footnote{Notice that we have changed the notation for the double Lorentz parameter $\Lambda \rightarrow \Gamma$.}. We dedicate the next part for doing so.  

\subsubsection{Extended symmetries and gauge fixing}
The symmetry transformations of the $O(D,D+K)$ fundamental fields are
\bea
\delta {\cal E}_{\cal M A} & = &  \hat \xi^{\cal N} \partial_{\cal N} {\cal E}_{\cal M A} + (\partial_{\cal M} \hat \xi^{\cal P} - \partial^{\cal P} \hat \xi_{\cal M}) {\cal E}_{\cal P A} + f_{\cal M N P} \hat \xi^{\cal N} {\cal E}^{\cal P}{}_{\cal A} + {\cal E}_{\cal M B} \Gamma^{\cal B}{}_{\cal A}  \, , \\
\delta d & = & \hat \xi^{\cal N} \partial_{\cal N} d - \frac{1}{2} \partial_{\cal M} \hat \xi^{\cal M} \, ,
\eea
where $f_{\cal MNP}$ only takes values when ${\cal MNP}=\alpha \beta \gamma$. The $\partial_{\cal M}$ derivative  is split according to $\partial_{\cal M}=(\partial_{M},0)$, while the generalized diffeomorphism parameter as 
\bea
\hat \xi^{\cal M} & = & (\hat{\xi}^{M}, \lambda^{\alpha}) \, .
\eea
Using the previous decomposition, $\delta d$ is what we expect. On the other hand, the double Lorentz parameter, ${\Gamma_{\cal AB}}$, requires a gauge fixing to ensure $\delta {\cal E}_{\alpha}{}^{\ov A} = 0$ and $\delta e_{i}{}^{\ov i}=0$. From the former we find,
\bea
\delta {\cal E}_{\alpha}{}^{\ov A} = -\partial^{P}\lambda_{\alpha} {\cal E}_{P}{}^{\ov A} + {\cal E}_{\alpha}{}^{\ov \b} \Gamma_{\ov \b}{}^{\ov A} = 0 \, .
\eea
Notice that ${\cal E}_{M \ov A}=E_{M \ov A}$, and then,
\bea
\Gamma_{\ov \b}{}^{\ov A} = (\Box^{-1/2})^{\alpha}{}_{\beta} e^{\beta}{}_{\ov \b} (\partial^{P} \lambda_{\alpha}) E_{P}{}^{\ov A} \, . 
\label{gfnew}
\eea
Similarly to the previous computation, we can now demand $\delta {\cal E}_{\alpha}{}^{\ov \alpha} = \delta(\Box^{\frac12})_{\alpha}{}^{\beta} e_{\beta}{}^{\ov \alpha}$, which is the condition that we need to impose $\delta e_{\a}{}^{\ov \a}=0$. This condition is satisfied if
\bea
\Gamma_{\ov \b \ov \a} = (\Box^{-1/2})^{\a}{}_{\b} e^{\b}{}_{[\ov \b} \Big(- \hat \xi_{P} \partial^{P}{\cal E}_{\alpha \ov \alpha]} + \partial^{P} \lambda_{\alpha} {\cal E}_{P \ov \a]}  - f_{\alpha \delta \gamma} \lambda^{\delta} {\cal E}^{\gamma}{}_{\ov \a]} + \delta_{\Lambda}({\Box^{\frac12}})_{\alpha \delta} e^{\delta}{}_{\ov \a]} \Big) \, .
\label{gfpar}
\eea
Then, equations (\ref{gfnew}) and (\ref{gfpar}) are the gauge fixing conditions that we need to write a GDFT in terms of fundamental fields which are in representations of $O(D,D)$. In the next part we will explore the transformation rules for these fields. 

\subsubsection{Transformation rules for the $O(D,D)$ field content}
We start by defining the following field
\bea
C_{M \alpha} = - A_{M}{}^{\beta} (\Box^{-\frac12})_{\beta \a}
\eea 
which is constrained by $E_{M}{}^{\ov A} C_{M \alpha}=0$. We also define
\bea
\Delta_{\alpha \beta} & = & \kappa_{\a \b} + C_{M \alpha} C^{M}{}_{\b} \\
\Theta_{MN} & = & \eta_{MN} + C_{M}{}^{\a} C_{N \a} \, .
\eea
Now we inspect the transformation of $\delta {\cal E}_{M \ov A}=\delta {E}_{M \ov A}$. In terms of the $C_{M \alpha}$ field we have,
\bea
\delta {E}_{M \ov A} = {\cal L}_{\hat \xi} E_{M \ov A} + E_{M}{}^{\ov B} \Lambda_{\ov B \ov A} + C_{M}{}^{\gamma} \partial^{P}\lambda_{\gamma} E_{P \ov A} \, ,
\eea
where we have identified
\bea
\Gamma_{\ov A \ov B} = \Lambda_{\ov A \ov B} \, .
\eea
At this point we observe that the transformation rule for $\delta E_{M \ov A}$ now contains an extra gauge transformation. Our idea is to construct gauge generators that allow us to identify
\bea
C_{M}{}^{\gamma} \partial_{P}\lambda_{\gamma} E^{P \ov A} \rightarrow \frac{a}{2} F_{\un M}{}^{\ov B \ov C} \partial_{P}\Lambda_{\ov B \ov C} E^{P \ov A} \, ,
\label{firstid}
\eea
similarly to the Bergshoeff-de Roo identification. Let's keep this idea on mind but, before defining the gauge generators at the DFT level, we need to inspect the other generalized frame projection $\delta E_{M \un A}$, 
\bea
\delta {\cal E}_{M \un A} = \delta(\chi^{\frac12})_{M}{}^{N} E_{N}{}^{\un A} + (\chi^{\frac12})_{M}{}^{N} \delta E_{N}{}^{\un A} \, .
\eea
From the previous expression we find,
\bea
\delta E_{N \un A} = {\cal L}_{\hat \xi} E_{N \un A} + E_{N}{}^{\un B} \Gamma_{\un B \un A} - (\Theta^{\frac12})^{M}{}_{N} \Big(\partial_{M} \lambda^{\a} C^{R}{}_{\beta} (\Delta^{\frac12})^{\beta}{}_{\a} + \delta_{\Lambda} (\Theta^{-\frac12})_{M}{}^{R} \Big) E_{R \un A} \, ,
\label{trframedown}
\eea
which need a parameter redefinition since we want (\ref{trframedown}) to have a generalized Green-Schwarz form. Then, we must define
\bea
Q_{N}{}^{R} = (\Theta^{\frac12})^{M}{}_{N} \Big( \partial_{M} \lambda^{\a} C^{R}{}_{\beta} (\Delta^{\frac12})^{\beta}{}_{\a} +  \delta_{\Lambda} (\Theta^{-\frac12})_{M}{}^{R} \Big) E_{R \un A} \, ,
\eea
and $S_{MN}=\ov P_{M}{}^{P} \partial_{P} \lambda^{\alpha} C_{N \alpha}-Q_{MN}$, in order to identify
\bea
\Gamma_{\un A \un B} = \Lambda_{\un A \un B} - E^{M}{}_{\un A} S_{MN} E^{N}{}_{\un B} \, .
\eea
In terms of the previous parameter, the transformation of this component takes its expected form,
\bea
\delta E_{N \un A} = {\cal L}_{\hat \xi} E_{N \un A} + E_{N}{}^{\un B} \Lambda_{\un B \un A} - \partial_{\ov N} \lambda^{\alpha} C^{R}{}_{\alpha} E_{R \un A} \, .
\label{frameun}
\eea
The only remaining transformation is $\delta C_{M \alpha}$. There exists several ways to obtain it. For instance, if we focus on $\delta {\cal E}_{M}{}^{\ov \a}$ we can obtain the transformation of the $C$-field (or $A$-field) according to
\bea
\delta {\cal E}_{M}{}^{\ov \a} = \delta ( - A_{M}{}^{\alpha}) e_{\alpha}{}^{\ov \alpha} = \delta (C_{M}{}^{\gamma} (\Box^{\frac12})_{\gamma}{}^{\beta}) e_{\beta}{}^{\ov \a} \, .
\eea

From the previous expression we find,
\bea
\delta C_{M}{}^{\gamma} = {\cal L}_{\hat \xi} C_{M}{}^{\gamma} + \partial_{M} \lambda^{\gamma} - (\Box^{-1})^{\gamma \alpha} \partial_{\ov M} \lambda_{\alpha} + C_{M}{}^{\delta} (\Box^{\frac12})_{\delta}{}^{\epsilon} e_{\epsilon}{}^{\ov \b} \Gamma_{\ov \b}{}^{\ov \a} e^{\b}{}_{\ov \a} (\Box^{-\frac12})^{\gamma}{}_{\beta} \, .
\eea
Moreover, using (\ref{gfpar}) one gets
\bea
\delta C_{M}{}^{\gamma} = {\cal L}_{\hat \xi} C_{M}{}^{\gamma} + \partial_{M} \lambda^{\gamma} - (\Box^{-1})^{\gamma \alpha} \partial_{\ov M} \lambda_{\alpha} + C_{M}{}^{\alpha} \partial^{P}\lambda_{\alpha} C_{P}{}^{\gamma} - C_{M}{}^{\alpha} f_{\alpha \beta}{}^{\gamma} \lambda^{\beta} \, . 
\label{Ctransf}
\eea
So far we have computed the transformation rules for all the fundamental fields of this formulation of gauged DFT. In the next part we will introduce the gauge generator $(t_{\alpha})^{\ov A \ov B}$ to obtain (\ref{firstid}). 

\subsubsection{First-order identification}
We start by taking the leading order contribution from the C-transformation (\ref{Ctransf}),
\bea
\delta C_{M}{}^{\gamma} = {\cal L}_{\hat \xi} C_{M}{}^{\gamma} + \partial_{\un M} \lambda^{\gamma} - C_{M}{}^{\alpha} f_{\alpha \beta}{}^{\gamma} \lambda^{\beta} \, . 
\label{Clead}
\eea
Indeed we are expecting that the second term of this transformation turns into a non-covariant double Lorentz transformation and the third one into a covariant double Lorentz transformation, as in (\ref{FluxM}). If this happens, we could identify the $C$-field with $F_{\un M \ov B \ov C}$. 

We define the gauge generators $(t_{\alpha})^{\ov A \ov B}$ such that,
\bea
(t^{\alpha})_{\ov A \ov B} (t_{\beta})^{\ov A \ov B} & = & X_{R} \delta_{\beta}^{\alpha} \, , \\
(t^{\alpha})_{\ov A \ov B} (t_{\alpha})^{\ov C \ov D} & = & X_{R} \delta_{\ov A \ov B}^{\ov C \ov D} \, ,
\label{mapDFT}
\eea
and for generic gauge/Lorentz vectors and parameters we have
\bea
V_{\ov B \ov C} & = & - V_{\alpha} (t^{\alpha})_{\ov B \ov C} \, , \\
\lambda_{\ov B \ov C} & = & - \lambda_{\alpha} (t^{\alpha})_{\ov B \ov C} \, .
\eea
The previous expressions mean that we are identifying the gauge symmetry of GDFT with the right sector of the double Lorentz transformations. Once these identifications are imposed, the GDFT turn into a DFT with higher-derivative corrections. 

The relation between the structure constants and the gauge generators is
\bea
f_{\alpha \beta}{}^{\gamma} (t_{\gamma})^{\ov A \ov B} = 2 (t_{[\alpha})^{[\ov A \ov C} (t_{\beta]})_{\ov C}{}^{\ov B]} \, .
\label{struct}
\eea
The map (\ref{mapDFT}) can be applied on (\ref{Clead}), 
\bea
\delta C_{M \ov B \ov C} = {\cal L}_{\hat \xi} C_{M \ov B \ov C} + \partial_{\un M} \lambda_{\ov B \ov C} + C_{M}{}^{\alpha} f_{\alpha \beta \gamma} \lambda^{\beta} (t^{\gamma})_{\ov B \ov C} \, ,
\eea
and using (\ref{struct}) one gets
\bea
\delta C_{M \ov B \ov C} = {\cal L}_{\hat \xi} C_{M \ov B \ov C} + \partial_{\un M} \lambda_{\ov B \ov C} + 2 C_{M [\ov B|\ov D} \lambda^{\ov D}{}_{\ov C]} \, .
\eea
Finally the identification is 
\bea
\label{idGS1}
C_{M \ov B \ov C} & = & F_{\un M \ov B \ov C} \, , \\ \lambda_{\ov B \ov C} & = & \Lambda_{\ov B \ov C} \, ,
\label{idGS2}
\eea
as expected. The correction to the transformation of the projections of the generalized frame now takes the form of a generalized Green-Schwarz mechanism,
\bea
\delta_{\Lambda} E_{M \ov A} & = & E_{M \ov B} \Lambda^{\ov B}{}_{\ov A} + \frac{1}{X_R} F_{\un M}{}^{\ov B \ov C} \partial^{P} \Lambda_{\ov B \ov C} E_{P \ov A} \, , \\
\delta_{\Lambda} E_{M \un A} & = & E_{M \un B} \Lambda^{\un B}{}_{\un A} - \frac{1}{X_R} \partial_{\ov M} \Lambda^{\ov B \ov C} F^{\un R}{}_{\ov B \ov C} E_{R \un A} \, ,
\eea
and therefore $\frac{1}{X_{R}}=\frac{a}{2}$. Up to this point, all our effort has produced the same corrections that we have proposed in the previous lecture to deform the Lorentz transformations. However, the identifications (\ref{idGS1}) and (\ref{idGS2}) can also be applied on the $O(D,D+K)$ action principle in order to construct the $4$-derivative corrections to the heterotic DFT Lagrangian. This method can be performed to obtain the corrections associated to the b-parameter, so we have found a systematic procedure to construct the bi-parametric first-order corrections to the DFT action. We show the form of all these corrections in the next part. 

\subsubsection{Fluxes and action principle}
In order to construct the higher-derivative action principle we need some components of the fluxes $\Big\{{\cal F}_{\cal A}, {\cal F}_{\cal A B C} \Big\}$. The $O(D,D+K)$ action is given by, 
\bea
\int d^{2D+K}X e^{-2d} {\cal R}({\cal E},d) \, ,
\eea
where ${\cal R}({\cal E},d)$ is,
\bea
{\cal R} & = & 2{\cal E}_{\cal \un{A}} {\cal F}^{\cal \un{A}} + {\cal F}_{\cal \un{ A}}{\cal F}^{\cal\un{A}} - \frac16{\cal F}_{\un{\cal ABC}}{\cal F}^{\cal \un{ABC}} - \frac12{\cal F}_{\ov{\cal A}\un{\cal BC}}{\cal F}^{\ov{\cal A}\un{\cal BC}} \, .
\label{act}
\eea
It turns out that the expression for the generalized fluxes is
\bea
{\cal F}_{\cal A} & = & \sqrt{2} \partial^{\cal M} {\cal E}_{\cal M A} - 2 {\cal E}_{\cal A} d \, , \\
{\cal F}_{\cal A B C} & = & 3 {\cal E}_{[\cal A} {\cal E}^{\cal M}{}_{\cal B} {\cal E}_{\cal M C]} + \sqrt{2} f_{\cal M N P} {\cal E}^{\cal M}{}_{\cal A} {\cal E}^{\cal N}{}_{\cal B} {\cal E}^{\cal P}{}_{\cal C} \, . 
\eea
Let us start inspecting the single index flux. The identification between $A_{M \alpha}$ and the generalized flux is 
\bea
A_{M \alpha} = - C_{M \alpha} \rightarrow - F_{\un M \ov B \ov C}
\eea
and the first order contribution to the generalized frame with underline index is
\bea
{\cal E}_{M \un A} = E_{M \un A} - \frac12 A_{M}{}^{\alpha} A_{N \alpha} E^{N}{}_{\un A} \, ,
\eea
according to (\ref{chiex}).
Using the previous expressions, the underline component of the single-index flux is
\bea
{\cal F}_{\un A} = \sqrt{2} \partial^{M}(E_{M \un A} - \frac12 A_{M}{}^{\alpha} A^{N}{}_{\alpha} E_{N \un A}) - 2 \sqrt{2} (E_{M \un A} - \frac12 A_{M}{}^{\alpha} A^{N}{}_{\alpha} E_{N \un A}) \partial^{M}d \, . 
\label{inter1}
\eea
We can rewrite (\ref{inter1}) as
\bea
{\cal F}_{\un A} = F_{\un A} - \frac14 F^{\un B} A_{\un B}{}^{\alpha} A_{\un A \alpha} - \frac14 E^{\un B}(A_{\un B}{}^{\alpha} A_{\un A \alpha}) \, 
\eea
using the notation $A_{\un B \alpha}= \sqrt{2} A_{M \alpha} E^{M}{}_{\un B}$. Now we impose the identification,
\bea
A_{\un B}{}^{\alpha} A_{\un A \alpha} = \frac{1}{X_{R}} A_{\un B}{}^{\ov C \ov D} A_{\un A \ov C \ov D} = \frac{a}{2} F_{\un B}{}^{\ov C \ov D} F_{\un A \ov C \ov D}  
\eea
to finally obtain
\bea
{\cal F}_{\un A} = F_{\un A} - \frac{a}{8} F^{\un B} F_{\un B}{}^{\ov C \ov D} F_{\un A \ov C \ov D} - \frac{a}{8} E^{\un B}(F_{\un B}{}^{\ov C \ov D} F_{\un A \ov C \ov D}) \, .
\eea

The same can be done for the different projections of ${\cal F}_{\cal A B C}$,
\bea
\label{genfluxes1}
{\cal F}_{\underline {ABC}}&=& F_{\underline {ABC}}  + \frac{3a}{4}\left(  E_{[\underline{A}}{F}^{\overline{CD}}{}_{\underline{B}} - 
\frac{1}{2}F_{\underline D[\underline{AB}}F^{\underline{D}\overline{CD}} - \frac{2}{3}F^{\overline{C}}{}_{\overline{E}[\underline{A}}F_{\underline{B}}{}^{\overline{ED}}\right)F_{\underline{C}]\overline{CD}}\, , \\ \label{genfluxes2}
{\cal F}_{\overline A\underline {BC}}&=& F_{\overline A\underline {BC}} +\frac{a}{4}\left(E_{\overline{A}}{F}^{\overline{CD}}{}_{[\underline{B}} +{F}^{\underline{E}\overline{CD}}{F}_{\overline{A}\underline{E}[\underline B}\right)F_{\underline{C}]\overline{CD}}\, ,    \\ \label{genfluxes3}
{\cal F}_{\underline A \overline {BC}} &=&  {F}_{\underline A \overline {BC}}-\frac{a}{8} F_{\underline D\overline {EF}} F^{\overline {E F}}{}_{\underline A} F^{\underline D }{}_{\overline {BC} }\, . 
\eea
If we now analyze the four-derivative contributions to the action (\ref{act}) we find,
\bea
2 {\cal E}^{\un A} {\cal F}_{\un A} & = & 2 {\cal E}^{\un A} {F}_{\un A} + 2 {E}^{\un A} {\cal F}_{\un A} \nn \\
{\cal F}_{\un A} {\cal F}^{\un A} & = & 2 {F}_{\un A} {\cal F}^{\un A} \nn \\
- \frac16 {\cal F}_{\un A \un B \un C} {\cal F}^{\un A \un B \un C} & = & - \frac13 {\cal F}_{\un A \un B \un C} {F}^{\un A \un B \un C} \nn \\
-\frac12 {\cal F}_{\ov {\cal A}  \un B \un C} {\cal F}^{\ov {\cal A} \un B \un C} & = & - { F}_{\ov {A}  \un B \un C} {\cal F}^{\ov {A} \un B \un C} - \frac12 {\cal F}_{\ov {\a}  \un B \un C} {\cal F}^{\ov {\a} \un B \un C}  \, . 
\label{actionterm}
\eea
The only flux projection that we have not computed yet is 
\bea
{\cal F}_{\ov {\a}  \un B \un C}  = - \frac{1}{\sqrt{2}} A^{\un D}{}_{\alpha} e^{\alpha}{}_{\ov \a} F_{\un D \un B \un C} + \frac12 f_{\alpha \beta \gamma} e^{\alpha}{}_{\ov \alpha} A_{\un B}{}^{\beta} A_{\un C}{}^{\gamma} \, , 
\eea
and replacing all the flux projections in (\ref{actionterm}) we get,
\bea
{\cal R}({\cal E},d) = {\cal R}(E,d) + a {\cal R}^{(-)} 
\eea
where
\bea
\label{R_minus}
{\cal R}^{(-)}& = & - \frac{1}{4}\left[(E_{\un{A}}E_{\un B}{F}^{\un B}{}_{\overline{CD}}) {F}^{\un{A}\overline{CD}} + (E_{\un{A}}E_{\un B}{F}^{\un A}{}_{\overline{CD}}) {F}^{\un{B}\overline{CD}}+ 2(E_{\un A}{F}_{\un B}{}^{\overline{CD}}){F}^{\un A}{}_{\overline{CD}}{F}^{\un{B}}\right.\, \\
& & + (E_{\un{A}}{F}^{\un{A}\overline{CD}})(E_{\un B}{F}^{\un B}{}_{\overline{CD}}) + (E_{\un A} {F}_{\un B}{}^{\overline{CD}})(E^{\un A} {F}^{\un B}{}_{\overline{CD}})+ 2(E_{\un{A}}F_{\un B}){F}^{\un B}{}_{\overline{CD}}{F}^{\un{A}\overline{CD}}\, \nn \\ 
& & + (E_{\overline{A}}F_{\un{B}\overline{CD}})F_{\un{C}}{}^{\overline{CD}}F^{\overline{A}\un{BC}} - (E_{\un{A}}F_{\un{B}\overline{CD}})F_{\un{C}}{}^{\overline{CD}}{F}^{\underline{ABC}} + 2(E_{\un A}{F}^{\un A}{}_{\overline{CD}}){F}_{\un B}{}^{\overline{CD}}{F}^{\un{B}}\, \nn \\
& &   - 4(E_{\un A} {F}_{\un B}{}^{\overline{CD}}){F}^{\un A}{}_{\overline{CE}} {F}^{\un B\overline{E}}{}_{\overline{D}}+ \frac{4}{3}{F}^{\overline{E}}{}_{{\un A}\overline{C} }{F}_{{\un B}\overline{ED}}F_{\un{C}}{}^{\overline{CD}}{F}^{\underline{ABC}} + {F}^{\un B}{}_{\overline{CD}} {F}_{\un A}{}^{\overline{CD}} F_{\un B}{F}^{\un{A}} \, \nn \\
& &  \left.+ {F}_{\un{A}}{}^{\overline{CE}}{F}_{\un B\overline{E}\overline{D}}{F}^{\un A}{}_{\overline{CG}} {F}^{\un B\overline{G}\overline{D}}
 - {F}_{\un{B}}{}^{\overline{CE}}{F}_{\un A\overline{E}\overline{D}}{F}^{\un A}{}_{\overline{CG}} {F}^{\un B\overline{G}\overline{D}} - F_{\overline{A}\underline{BD}}F^{\un{D}}{}_{\overline{CD}}F_{\un{C}}{}^{\overline{CD}}F^{\overline{A}\un{BC}}\right]\, . \nn
\eea
The systematic procedure used to construct ${\cal R}^{(-)}$ can be easily adapted to construct ${\cal R}^{(+)}$. The strategy that one should follow is to start from an $O(D+K,D)$ invariant theory and to identify
\bea
C_{M \alpha} = C_{\ov M \alpha}
\eea
with $F_{\ov M \un B \un C}$ considering gauge generators that allow the index identification $\alpha \rightarrow [\un B \un C]$. This other correction to ${\cal R}(E,d)$, that we will call it ${\cal R}^{(+)}$, has the form (\ref{R_minus}) but exchanging the projections of the different fields. Then, the action principle
\bea
S = \int d^{2D}X e^{-2d} ({\cal R}(E,d) + a {\cal R}^{(-)} + b {\cal R}^{(+)}) \, ,
\label{final}
\eea
is the explicit form of (\ref{actionfull}). The invariance of this action under double Lorentz transformations was studied in \cite{Tduality} considering an hybrid formalism that mixes the generalized flux formalism with the generalized metric formalism. The equivalence between (\ref{final}) and the hybrid formulation was given in \cite{Odd}.

\subsection{Part B: extension to higher-order formulations}
\label{B3}
In the previous part we used a systematic procedure to deform both the symmetries and the action of DFT. The double Lorentz transformations are deformed by a generalized Green-Schwarz mechanism that was constructed considering an $O(D,D+K)$ (or $O(D+K,D)$) invariant theory and we split its fundamental fields in terms of $O(D,D)$ multiplets. Using this method, we obtained a gauged DFT with generalized gauge fields $C_{M \alpha}$ which we identified with a suitable projection of the generalized fluxes. However, this procedure cannot be extended to the next order using (\ref{idGS1}) since the gauge field does not transform as a flux projection. This problem was solved in \cite{gbdr}. We explain some details of the resolution in the following part.

\subsubsection{The generalized Bergshoeff-de Roo identification}
In \cite{gbdr} the authors proposed to identify the gauge group of an $O(D,D+K)$ theory with $O(1,D+K-1)_{R}$. At first sight, this would seem impossible since the dimension of both groups does not match for finite $K$, \textit{i.e.},
\be
K = \frac {(D+K)(D+K-1)}{2} \ ,
\label{Dimension}
\ee
However, the identification works when $K \to \infty$ \footnote{This method is heuristic. Its mathematical foundation deserve further study, as the authors commented in \cite{gbdr}.}. First we need to introduce the map
\be
 V_{\overline {\cal A}}{}^{\overline {\cal B}} = - \, V_\alpha \, (t^\alpha)_{\overline {\cal A}}{}^{\overline {\cal B}} \ , \label{IndexRelation}
\ee
where $\left(t_\alpha\right)_{\overline{\cal A}}{}^{\overline{\cal B}}$ denote the generators of the gauge algebra. We impose ${A}_{\overline M \alpha} = 0$ and $e_\alpha{}^{\overline \alpha} = \rm{const}.$, which require non vanishing $\Gamma_{\overline {A \alpha}}$ and $\Gamma_{\overline{\alpha \beta}}$ parameters. The only degree of freedom to identify is 
\bea
{\cal E}_{\alpha \underline A} = E^M{}_{\underline A} {A}_{M\alpha} = \frac{1}{\sqrt{2}} {A}_{\underline A \alpha} \, ,  
\eea
which transforms as
\be
\delta {A}_{\underline A \alpha} = \hat \xi^{P} \partial_{P} A_{\un A \alpha} - {\cal E}_{\underline A} \lambda_\alpha + f_{\alpha \beta}{}^\gamma \lambda^\beta {A}_{\underline A \gamma} + { A}_{\underline B \alpha} \Gamma^{\underline B}{}_{\underline {A}} \ . \label{TransformationA}
\ee
Using the map \eqref{IndexRelation} and the relation between the structure constants and the generators
\bea
f_{\alpha \beta}{}^{\gamma} (t_{\gamma})^{\ov {\cal A} \ov {\cal B}} = 2 (t_{[\alpha})^{\ov {\cal A} \ov {\cal C}} (t_{\beta]})_{\ov {\cal C}}{}^{\ov {\cal B}} \, ,
\eea
(\ref{TransformationA}) can be written as
\begin{eqnarray}
 \delta {A}_{\underline A \overline{\cal B C}}= {\cal L}_{\hat \xi} {A}_{\underline A \overline{\cal B C}} - {\cal E}_{\underline A} \lambda_{\overline{\cal B C}} +  2 { A}_{\underline A \overline{\cal D} [\overline{\cal C}}\, \lambda^{\overline{\cal D}}{}_{\overline{\cal B}]} +  {A}_{\underline D \overline{\cal B C}}\, \Gamma^{\underline{D}}{}_{\underline A} \ . \label{TransformationAH}
\end{eqnarray}
This is precisely the way in which the following projection of the extended fluxes transforms
\bea
\delta {\cal F}_{\underline A \overline {\cal B C}} = {\cal L}_{\hat \xi} {\cal F}_{\underline A \overline {\cal B C}} + {\cal E}_{\underline A} \Gamma_{\overline{\cal B C}} + 2 {\cal F}_{\underline A  \overline {\cal D} [\overline{\cal C}} \Gamma^{\overline{\cal D}}{}_{\overline{\cal B}]} + {\cal F}_{\underline D  \overline {\cal B C}} \Gamma^{\underline D}{}_{\underline {A}} \ . \label{TransformationFProjected}
\eea

The resemblance between \eqref{TransformationAH} and \eqref{TransformationFProjected} turns into an exact identification provided by
\bea
\lambda_{\overline {\cal A B}} & = & -  \, \lambda_\alpha \, (t^\alpha)_{\overline {\cal A B}} \ =\ \Gamma_{\overline {\cal A B}}\ , \\
{A}_{\underline A \overline {\cal B C}} & = & -  \, {\cal E}_{\alpha \underline A} \, (t^\alpha)_{\overline {\cal B C}} \ =\ - {\cal F}_{\underline A \overline {\cal B C}}  \ .
\label{genbdr}
\eea
In \cite{gbdr} these identifications were named as the generalized Bergshoeff-de Roo identification. Interesting enough, this recursive procedure is useful to construct higher-order contributions, where $X_R$ acts as a regulator. The method can be implemented in an iterative way, following the rule
\bea
{A}_{\un A \cal \ov B \ov C } \rightarrow -{\cal F}_{\un A {\cal \ov B \ov C}}({A}_{\un A \cal \ov B \ov C } \rightarrow -{\cal F}_{\un A {\cal \ov B \ov C}}({A}_{\un A \cal \ov B \ov C } \rightarrow -{\cal F}_{\un A {\cal \ov B \ov C}}(\dots)))  \, ,
\eea
and converges up to second-order (and possible to all orders.) The state of the art related to this procedure is discussed in the next part.

\subsubsection{Current research}

The generalized Bergshoeff-de Roo identification discussed in the last part of these lectures contains a supersymmetric extension that we have not reviewed. In \cite{LNR} the authors showed how this extension works to first-order in $\alpha'$. Therefore, an explicit relation between the four-derivative heterotic supergravity and a higher-derivative
${\cal N} = 1$ supersymmetric DFT was achieved. Moreover, since the identification (\ref{genbdr}) is exact, the procedure can be performed to obtain corrections to any desired order.  However, it is likely that higher-order corrections do not reproduce the full string theory Lagrangian. The reason of the previous statement is related to quartic Riemann interactions, which contain terms proportional to the transcendental coefficient $\zeta(3)$ in heterotic string theory  \cite{GrossSloan} \cite{CN}. These kind of contributions presumably require additional deformations to the generalized Bergshoeff-de Roo identification. Other possibility is the existence of a new invariant in DFT, with eight derivatives. This means that our starting point could be a higher derivative GDFT, and then the iterative procedure introduce new corrections. Nevertheless, it was recently shown in \cite{Wulff} that such an object cannot be constructed in a T-duality invariant approach before compactification, so this interesting problem is still open. 

In \cite{gbdr2}, the extension to include second-order corrections beyond heterotic DFT/supergravity was achieved. The idea of the authors was to start with an extended duality group $O(D + p,D + q)$ and then perform an $O(D,D)$ decomposition. This method is an extension of the generalized Bergshoeff-de Roo identification that captures the bi-parametric freedom, but the quartic Riemann interaction problem is still present. On the other hand, higher-derivative formulations have been recently used at the level of perturbative DFT to inspect double copy relations from a duality covariant approach \cite{AleyEric2}, in non-Abelian and Poisson-Lie T-duality \cite{Poisson} and in closely related integrable deformations of the string sigma model \cite{Integrable}. The study of higher-dimensional black holes and entropy was also studied considering higher-derivative deformations \cite{black}. Finally, several works restrict the study of the higher-derivative corrections to backgrounds with commuting isometries \cite{iso}.

\newpage
\subsection{Exercises}
\begin{enumerate}
    \item Compute the leading order and the next-to-leading order contributions to $\Gamma_{\ov \a \ov B}$ and $\Gamma_{\ov a \ov b}$. 
    
    \item * Show that $\Gamma_{(\ov a \ov b)}=0$, and therefore expression (\ref{gfpar}) holds.
    
    \item Prove the following identities: $\Delta_{\a \b} = (\Box^{-1})_{\a \b}$ and $\chi_{M N} = (\Theta^{-1})_{MN}$ \, .

\item Starting from $\delta {\cal E}_{M \un A}$, show that (\ref{frameun}) holds.

\item * Show that $S_{(MN)}=0$ and $S_{MN}=S_{\un M \un N}$.

\item Starting from $\delta {\cal E}_{M \ov \a}$, show that (\ref{Ctransf}) holds.

\item Compute the next-to-leading order correction to (\ref{Clead}). Conclude that the identification (\ref{idGS1}) is not valid to next order.

\item Compute the flux components (\ref{genfluxes1})-(\ref{genfluxes3}).

\item Compute ${\cal R}^{(-)}$ and show that (\ref{R_minus}) holds.

\item Show that there is no solution to (\ref{Dimension}) for finite $K>1$.

\item * Show that the generalized Bergshoeff-de Roo identification is compatible to first-order finding $X_R$ in terms of the $a$ parameter.

\end{enumerate}
\newpage
\section{Summary}
In these lectures we have reviewed some aspects of the low-energy limit of string theory when four-derivative contributions are considered. We focus on the gravitational corrections for the different formulations of closed string theory. We have studied the Metsaev-Tseytlin approach, related to scattering amplitude computations, and the Bergshoeff-de Roo approach, which can be obtained considering field redefinitions in the Metsaev-Tseytlin approach. In the Bergshoeff-de Roo approach, a relation between non-Abelian gauge symmetries and Lorentz transformations was analyzed for the heterotic supergravity. It turns out that the non-Abelian gauge transformations can be written as higher-derivative Lorentz transformations using gauge generators that relate gauge/Lorentz indices. The Green-Schwarz mechanism corrects only the $b$-field Lorentz transformation, since no other field different from the gauge field $A_{\mu i}$, transforms under non-Abelian gauge transformations. Interesting enough, the gravitational four-derivative terms in the heterotic supergravity action can be reproduced using this identification.

The generalization of the previous relation to the DFT level allowed us to propose a generalized Green-Schwarz mechanism to deform the double Lorentz symmetry for the generalized frame using a pair of parameters $(a,b)$. All choices reproduce T-duality invariant theories and only a few cases reproduce string supergravities. Among the non-string theories we discuss about the HSZ theory, which is a deformation of DFT that can be expressed in the generalized metric formulation. In this case, the generalized diffeomorphism transformations contain higher-derivative deformations, and the C-bracket is consistently deformed. At the supergravity Lagrangian level, the higher-derivative corrections of HSZ theory are closely related to heterotic string theory for odd powers of $\alpha'$ and to bosonic string theory for even powers of $\alpha'$.         

Finally, in the last lecture, we introduce a systematic procedure to construct deformations in DFT. The idea is to consider an $O(D,D+K)$ invariant theory and to write the fields, parameters and metrics in terms of $O(D,D)$ degrees of freedom. This procedure requires a suitable gauge fixing. The fundamental fields, from an $O(D,D)$ perspective, include a gauge field $C_{M \alpha}$ that can be identified with a projection of the generalized fluxes. Using this method, to first order, one can construct deformations for both the symmetries and the action principle in a systematic way. The generalization of this procedure to higher-orders (the generalized Bergshoeff-de Roo identification) was partially discussed together with some current research.

\section{Acknowledgments}

These notes were written for the lectures delivered at the school
“Integrability, Dualities and Deformations", that ran from 23 to 27 August
2021 in Santiago de Compostela and virtually." Website:
https://indico.cern.ch/e/IDD2021

The author is very grateful to W. Baron, D. Marques, C. Nunez and A. Rodriguez for collaboration in some works covered by these lectures. I warmly thank R. Borsato, E. Malek and D. Marques for enlightening comments which help me to organize the content of each lecture. I also indebted with S. Iguri, N. Miron-Granese and A. Rodriguez for helping me to improve the presentation of the first version of these notes, and with R. Borsato and E.Malek for helping me to improve the present version. Finally, I would like to thank Riccardo Borsato, Saskia Demulder, Sibylle Driezen, Fedor Levkovich-Maslyuk and Emanuel Malek for all the support and help during the school.

Support by CONICET is also gratefully acknowledged.

\end{document}